\documentclass{elsarticle}
\usepackage[utf8]{inputenc}
\usepackage[T1]{fontenc}  
\usepackage{lmodern}    
\usepackage[section]{placeins}
\usepackage{textcomp}
\usepackage{gensymb}
\usepackage{amsfonts}
\usepackage{amssymb}
\usepackage{amsmath}
\usepackage{glossaries}
\usepackage{siunitx}
\usepackage{subcaption}
\usepackage{hyperref}
\usepackage{multirow}
\usepackage{tabularx}
\usepackage{listings}
\usepackage{xcolor}

\def\equationautorefname~#1\null{eq. (#1)\null}
\DeclareMathOperator*{\argmin}{arg\,min}

\lstdefinestyle{plainoutput}{
  basicstyle=\ttfamily\footnotesize,
  numbers=left,
  numberstyle=\tiny,
  stepnumber=1,
  breaklines=true,          
  showstringspaces=false,
  keywordstyle=,
  commentstyle=,
  stringstyle=,
}

\newacronym{EVA}{EVA}{extreme value analysis}
\newacronym{PDF}{PDF}{probability density function}
\newacronym{CDF}{CDF}{cumulative distribution function}
\newacronym{GEV}{GEV}{generalized extreme value distribution}
\newacronym{GPD}{GPD}{generalized Pareto distribution}
\newacronym{KL}{KL}{Kullback–Leibler divergence}
\newacronym{MLE}{MLE}{maximum likelihood estimate}
\newacronym{CART}{CART}{classification and regression trees}
\newacronym{PWM}{PWM}{probability weighted moments}
\newacronym{iid}{i.i.d.}{independently and identically distributed}
\newacronym{DSM}{DSM}{demand-side management}
\newacronym{VaR}{VaR}{value-at-risk}
\newacronym{CVaR}{CVaR}{conditional value-at-risk}
\newacronym{EUE}{EUE}{expected unserved energy}
\newacronym{LOLP}{LOLP}{loss-of-load probability}
\newacronym{NERC}{NERC}{North American Electric Reliability Corporation}
\newacronym{CNN}{CNN}{convolution neural network}
\newacronym{QTCN}{QTCN}{quantile temporal convolution neural network}
\newacronym{NGD}{NGD}{natural gradient descent}
\newacronym{CRB}{CRB}{Cram\'er-Rao bound}
\newacronym{DOA}{DOA}{domain of attraction}
\newacronym{QRF}{QRF}{quantile regression forest}
\newacronym{LPHC}{LPHC}{low-probability high-consequence}
\newacronym{PPM}{PPM}{parts per million}
\newacronym{AI}{AI}{artificial intelligence}
\newacronym{HIA}{HIA}{hourly integrated average}
\newacronym{ML}{ML}{machine learning}
\newacronym{QR}{QR}{quantile regression}
\newacronym{ANN}{ANN}{artificial neural network}
\newacronym{KDE}{KDE}{kernel density estimation}
\newacronym{QF}{QF}{quantile function}
\newacronym{ACF}{ACF}{autocorrelation function}
\newacronym{MSE}{MSE}{mean squared error}
\newacronym{MVUE}{MVUE}{minimum-variance unbiased estimator}
\newacronym{RTO}{RTO}{regional transmission organization}
\newacronym{SCED}{SCED}{security-constrained economic dispatch}
\newacronym{LMP}{LMP}{locational marginal price}
\newacronym{MCP}{MCP}{market clearing price}
\newacronym{RAC}{RAC}{reliability assessment and commitment}
\newacronym{DASC}{DASC}{day-ahead scheduling capacity}
\newacronym{MAE}{MAE}{mean absolute error}
\newacronym{FOR}{FOR}{forced outage rate}
\newacronym{DAEM}{DAEM}{day-ahead energy market}
\newacronym{GAMLSS}{GAMLSS}{generalized additive model for location, scale and shape}
\newacronym{LOLE}{LOLE}{loss-of-load expectation}
\newacronym{AEMO}{AEMO}{Australian Energy Market Operator}
\newacronym{LSTM}{LSTM}{long short-term memory}
\newacronym{GARCH}{GARCH}{generalized autoregressive conditional heteroskedasticity}
\newacronym{USE}{USE}{unserved energy}
\newacronym{EENS}{EENS}{expected energy not served}
\newacronym{CRPS}{CRPS}{continuous ranked probability score}
\newacronym{ELM}{ELM}{extreme learning machines}

\begin{document}
    \title{Ensemble-Based Peak Demand Probability Density Forecasting with Application to Risk-Aware Power System Scheduling\tnoteref{funding}\tnoteref{source-code}}
    \tnotetext[funding]{This paper is partly derived from the work supported by the U.S. Department of Energy’s Office of Energy Efficiency and Renewable Energy (EERE) through the Solar Energy Technology Office (SETO) under Award DE-EE0009357. The funding source has no involvement in the development of the methodology, design, preparation, or interpretation of the case study, or the decision to submit for publication. }
    \tnotetext[source-code]{Source code artifacts are available in repository \url{https://github.com/byu9/extreme-tree}. Results in the paper correspond to revision \texttt{888ef5a}.}
    \author[ncsu]{Buyi Yu}\corref{cor1}\ead{byu9@ncsu.edu}
    \author[ncsu]{Wenyuan Tang}\ead{wtang8@ncsu.edu}
    \cortext[cor1]{Corresponding author}
    \affiliation[ncsu]{
        addressline={Campus Box 7571},
        organization={North Carolina State University},
        city={Raleigh, NC 27695},
        country={United States}
    }
    \begin{abstract}
Power systems face increasing challenges in maintaining resource adequacy due to lower operating margins, rising renewable energy uncertainty, and demand variability. Forecasting the probability distribution of peak demand on shorter timescales is a critical forward-facing issue under increasing volatility. This study introduces a novel ensemble-based machine learning method for peak demand probability density forecasting that extends classical extreme value theory to model time series peaks as nonstationary statistical distributions. The approach employs an ensemble of tree-based learners that recursively partition the covariate space and estimate local generalized extreme value distributions, allowing it to automatically capture complex covariate-dependent parameter variations. Unlike existing approaches, which often suffer from convergence issues or restrictive functional forms, this framework is both flexible and robust. Validation on a case study based on the PJM interconnection demonstrates that the method achieves a 38 percent reduction in committed capacity when generation is scheduled based on a reliability criterion. These improvements provide practical value for power system operation, enabling risk-aware capacity scheduling under peak demand uncertainty and supporting reliability-driven decision making in future energy systems.
\end{abstract}


    \maketitle
    \section{Introduction}
Increasingly frequent extreme weather events, such as the winter storm that resulted in the 2021 Texas power crisis, can disrupt power balance, leading to cascading failures of the power grid with devastating and sustained impacts \cite{flores_2021_2023}. 
Energy regulatory and oversight agencies worldwide, such as \gls{NERC}, \gls{AEMO}, and the European Union, enforce reliability requirements in the form of probabilistic measures formulated on the statistical behavior of peak demand. 
Such measures include the \gls{LOLP} in \gls{NERC} standard BAL-502-RF-03 \cite{north_american_electric_reliability_corporation_bal-502-rf-03_2025}, the \gls{EUE} in the Australian National Electricity Law \cite{australian_energy_market_commission_national_2025}, and the \gls{LOLE} in European regulation EU 2019/941 \cite{european_parliament_council_of_the_european_union_regulation_2025}.
These standards require resource allocation to balance operational cost against risk acceptability and account for the low-probability high-impact ``black swan'' events based on the worst-case characterization of demand uncertainty. 
Reflecting compliance with these standards, system operators often apply Mont\'e-Carlo simulations in annual studies to identify extreme demand scenarios that could result in severe energy shortfalls, such as in the ISO New England 2027 Winter Study \cite{iso_new_england_operational_2025}. 

Accurate characterization of peak demand under shorter intervals has become increasingly important under the growing volatility introduced by renewable generation and rapidly changing net-demand patterns.
Not only is such modeling a core element of maintaining system reliability, flexibility, and resilience of future energy systems, but it also enables secure real-time decision making and effective capacity planning under uncertainty.
A suitable framework for addressing this challenge must explicitly represent the statistical behavior of extreme demand events while accommodating nonstationary system conditions.
\Gls{EVA} provides a rigorous statistical foundation for modeling the probability of rare events \cite{bousquet_extreme_2021}.
Classical \gls{EVA} demonstrates that the block maxima of random observations converge to the \gls{GEV} under broad conditions, but it assumes stationarity, limiting its use in operational contexts where the distribution parameters vary with external factors.
In practice, \gls{EVA} is often combined with Mont\'e-Carlo simulation to overcome stationarity through case-by-case simulations.
However, the daily peak demand is influenced by multiple interacting factors, including local weather dynamics, seasonal and calendar effects, and demand patterns shaped by preceding days.
In such short-term settings, high-dimensional Mont\'e-Carlo simulation becomes computationally prohibitive.
This limitation highlights a critical methodological gap, underscoring the need for novel approaches that can flexibly model nonstationary peak demand distributions.

\subsection{Challenges with state-of-the-art probability density forecasting approaches}
Recent studies have emphasized the need for integrating probabilistic forecasting and risk management in renewable-based systems. For instance, \cite{hassan_integrated_2024} presents an integrated smart risk management framework combining forecasting, operational assessment, and decision strategies for renewable operation.
Probabilistic regression is the main approach to probabilistic forecasting and estimates likelihoods from deterministic observations. It can be broadly classified into the following two categories:

\paragraph{Parametric approaches}
In parametric probabilistic regression, the probable outcome $y(\boldsymbol{x})$ under the condition known as the covariate $\boldsymbol{x}$ is assumed to follow a predetermined distribution family $D(\hat{\boldsymbol\theta}(\boldsymbol {x}))$ with predicted deterministic parameter $\hat{\boldsymbol \theta}(\boldsymbol{x})$. 
Classical models such as \gls{GARCH} \cite{bollerslev_generalized_1986} capture only the nonstationarity in variance and are thus limited in flexibility.
A broader alternative, \gls{GAMLSS} \cite{rigby_generalized_2005}, models the nonstationary properties of three distribution parameters using specified functional forms. When combined with \gls{GEV}, \gls{GAMLSS} often suffers from convergence problems, as observed by \cite{debele_around_2017} in flood prediction and in the case study in \autoref{sec:case_study}.

\paragraph{Non-parametric approaches}
Non-parametric methods avoid fixed distribution families, instead estimating empirical quantiles, \gls{CDF}, or histograms. 
Examples include the well-known \gls{QR} \cite{koenker_regression_1978} and many deep learning methods that produce quantile forecasts.
State-of-the-art works include attention-based quantile methods for load forecasting \cite{guo_probabilistic_2024} and reinforcement-learning-based quantile models for wind and energy price \cite{li_probabilistic_2024}.
However, as it is demonstrated in \autoref{sec:example} and \autoref{sec:case_study}, quantile methods are poorly calibrated for extremes and cannot accurately approximate the probability density of rare ``black swan'' peak observations critical to system operations.

\paragraph{Review of cutting-edge literature}
Recent works on probabilistic forecasting are summarized in \autoref{tab:recent_works}, which also includes renewable energy forecasting literature since renewable generation is often modeled as negative demand.
The majority of recent literature focuses on predicting the probability density of the ordinary demand at a future time interval, and we refer to them as ``ordinary'' methods.
While these models provide useful insights into the statistical behavior of typical demand observations, which are concentrated in the bulk of the predicted density, they are fundamentally limited when used to forecast the probability density of peak demand observations, which are located at the tail of the predicted density.
As time series are often dominated by ordinary observations rather than extreme observations, regression accuracy is biased toward the bulk rather than the tail, where the system risks are the most critical. 
Consequently, ``ordinary'' density models are unsuitable for reliability assessment and risk-aware scheduling, highlighting the need for an \gls{EVA}-based framework that explicitly characterizes the tail behavior. 

\begin{table}[htb!]
    \centering
    \scalebox{0.75}{
    \begin{tabular}{llll}
        \hline
        Method &
        Year &
        Use Case &
        Classification
        \\
        \hline
        Conformal prediction \cite{zuege_wind_2025} &
        2025 & 
        (Ordinary) wind speed &
        Non-parametric, intervals
        \\
        Model screening \cite{parejo_probabilistic_2025} &
        2025 &
        (Ordinary) demand/generation &
        Parametric / non-parametric
        \\
        Nearest neighbor \cite{botman_global_2025} &
        2025 &
        (Ordinary) demand &
        Non-parametric, quantiles
        \\
        Reinforcement learning \cite{li_probabilistic_2024} &
        2024 &
        (Ordinary) wind power/price &
        Non-parametric, quantiles
        \\
        Transformer / GP \cite{hu_probabilistic_2024} &
        2024 &
        (Ordinary) demand &
        Parametric, GP
        \\
        Attentive QTCN \cite{guo_probabilistic_2024} &
        2024 &
        (Ordinary) demand &
        Non-parametric, quantiles
        \\
        Novel encoding \cite{zhu_peak_2024} &
        2024 &
        Extreme wind power & 
        Non-parametric, intervals
        \\
        Mont\'e-Carlo \cite{marulanda_modeling_2024} &
        2024 &
        (Ordinary) wind power & 
        Parametric, various
        \\
        Transformer \cite{xu_interpretable_2024} &
        2024 &
        (Ordinary) demand &
        Non-parametric, quantiles
        \\
        Deep learning \cite{zhu_ultra-short-term_2024} &
        2024 &
        (Ordinary) wind power &
        Non-parametric, quantiles
        \\
        Mont\'e-Carlo \cite{selcen_ayaz_probabilistic_2024} &
        2024 &
        (Ordinary) PV / demand &
        Non-parametric, histogram
        \\
        Improved TFT \cite{li_probabilistic_2023} &
        2023 &
        (Ordinary) demand &
        Non-parametric, quantiles
        \\
        Stacking ensemble \cite{he_short-term_2022} &
        2022 &
        (Ordinary) demand &
        Non-parametric, quantiles
        \\
        Mont\'e-Carlo \cite{gruosso_probabilistic_2019} &
        2019 &
        (Ordinary) demand &
        Parametric, normal 
        \\
        \hline
    \end{tabular}
    }
    \caption{Recent works on probabilistic forecasting, use case, and classification. GP--- Gaussian process, QTCN---quantile temporal convolution network, TFT---temporal fusion transformer.}
    \label{tab:recent_works}
\end{table}

Out of the reviewed literature, only \cite{zhu_peak_2024} specifically concentrates on the prediction of the peak density, where the uncertainty in peak wind power distribution is characterized using an interval model via a novel neural network encoding.
Interval forecasts are a special case of the quantile method where only a pair of quantile predictions is given, and thus fall under the non-parametric category. 
Non-parametric quantile methods are favored due to their flexibility and assumption-free nature, but have several notable shortcomings.
They tend to suffer from quantile crossover \cite{waldmann_quantile_2018}, where the approximated inverse \gls{CDF} is not monotonically non-decreasing, implying unsubstantial negative densities. 
Moreover, they produce only discrete estimates at selected quantiles, and to obtain a continuous density, post-processing such as \gls{KDE} \cite{rosenblatt_remarks_1956} is required to interpolate the missing quantiles.
While adequate for some applications, these drawbacks limit the usability of quantile-based peak demand density prediction for high-stakes risk management.

Key differences that distinguish the proposed approach from current literature are shown in \autoref{tab:distinguishing_features_of_our_method}.
The proposed approach aims to estimate the conditional distribution of peak demand using an \gls{EVA} framework and a novel tree-based ensemble. 
It is important to point out that, while being lexically similar, it is fundamentally different from the \gls{ELM} in  \cite{ertugrul_forecasting_2016}, in which the word ``extreme'' refers to its extremely fast training phase rather than the modeling of extreme values. Therefore, \cite{ertugrul_forecasting_2016} is considered a conventional point forecasting method, which is excluded from the literature review.

\begin{table}[htb!]
    \centering
    \scalebox{0.9}{
    \begin{tabular}{lll}
        \hline 
         Aspect & Existing methods & Proposed method \\
        \hline
        Prediction format & Series of mean or quantiles & Series of probability densities \\
        Forecasting target & Ordinary and peak values &  Only peak densities \\
        Accuracy of peaks & Often overlooked & Specifically targeted \\
        Assumptions & Often empirical & Conditional EVA framework \\
        Reliability metrics & Often limited to quantiles & VaR, CVaR, EUE, LOLP \\
        \hline
    \end{tabular}
    }
    \caption{Key differences that distinguish the proposed approach from existing forecasting literature. EVA---extreme value analysis, VaR---value-at-risk, CVaR---conditional value-at-risk, EUE---expected unserved energy, LOLP---loss-of-load probability.}
    \label{tab:distinguishing_features_of_our_method}
\end{table}

\subsection{Key contributions and paper structure}

The proposed approach is classified under the parametric category and utilizes a novel tree-based bagging ensemble for parameter estimation. This paper makes the following key contributions:

\begin{itemize}
  \item We formulate a statistical framework for modeling nonstationary peak demand densities, providing a rigorous basis for incorporating covariate effects into \gls{EVA}. 
  \item We develop a novel nonstationary method for estimating \gls{GEV} parameters under this framework, addressing the limitations of classical \gls{EVA} in stationarity and \gls{GAMLSS} in convergence.
  \item We validate the proposed method through an experiment against benchmark estimators, demonstrating improved performance in modeling extremes. 
  \item We apply the method in a real-world case study, showing how it can enhance industry practices by improving risk management and reducing operational costs. 
\end{itemize}

The remainder of the paper is organized as follows. In the next section, we describe the problem, present the conditional framework for nonstationary extrema, outline assumptions, and introduce two configurations for modeling autocorrelation. In \autoref{sec:methodology}, we detail the proposed nonstationary \gls{EVA} method, compare estimation strategies, and present the algorithm. \autoref{sec:example} reports an experiment comparing the proposed approach to parametric and nonparametric competitors, and \autoref{sec:case_study} demonstrates its practical benefits through a case study.

\subsection{Research workflow}

\begin{figure}[htb!]
    \centering
    \includegraphics[width=0.7\textwidth]{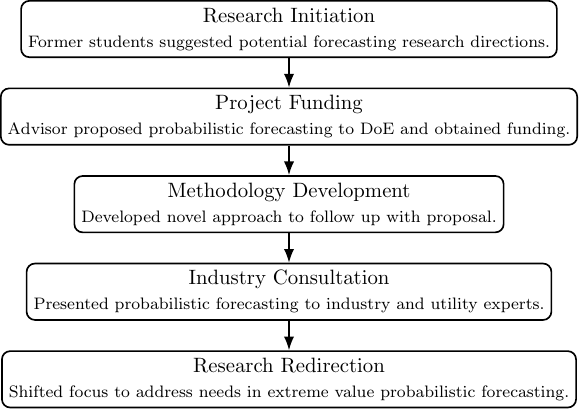}
    \caption{Research workflow. DoE---Department of Energy.}
    \label{fig:research_workflow}
\end{figure}

The research workflow is shown in \autoref{fig:research_workflow}. 
Early investigations into forecasting directions were suggested by former students, while subsequent feedback from industry and utility experts motivated a redirection of the research scope. 
Industry stakeholders indicated that improving point forecast accuracy is no longer their primary concern, since existing approaches already meet operational needs. 
Rather than pursuing marginal accuracy improvements in point forecasts, where deep-learning methods such as \gls{LSTM} \cite{hochreiter_long_1997} and temporal convolutional networks \cite{bai_empirical_2018} are already well established, this work focuses on developing a novel methodological contribution centered on extreme demand probabilistic modeling for risk-aware system operations.
The integration of deep learning with the proposed approach is viewed as an interesting direction for future exploration, but not as a necessary condition for establishing the value and practicality in its current form. 

    \section{Problem statement}
\label{sec:problem_statement}
\begin{figure}[htb!]
    \centering
    \includegraphics[width=0.8\textwidth]{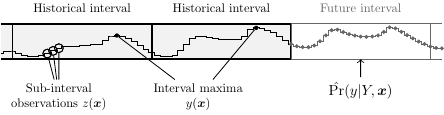}
    \caption{Sub-interval observations, interval maxima, and the predicted conditional probability density of the interval maximum.}
    \label{fig:model_interval}
\end{figure}
The phrase ``peak demand'' implies that it is an observation within a time interval.
For deterministic sub-interval condition $\boldsymbol{x}$, we use the symbol $Y$ to represent the event that the peak demand is observed at $\boldsymbol{x}$. Given the knowledge that event $Y$ occurs, the goal is to predict the distribution of the peak demand $\Pr(y|Y, \boldsymbol{x})$ based on future environmental conditions $\boldsymbol{x}$, which is also known and deterministic.  The relationship between sub-interval observations, intervals, and interval maxima is illustrated in \autoref{fig:model_interval}. Using $\Pr(Y|\boldsymbol{x})$ to represent the probability of sub-interval condition $\boldsymbol{x}$ resulting in peak demand event $Y$, the joint probability $\Pr(y, Y | \boldsymbol{x})$ of peak demand event $Y$ of severity $y$ occurring under sub-interval condition $\boldsymbol{x}$ is given by
\begin{equation}
    \Pr(y, Y | \boldsymbol{x}) = \Pr(y|Y, \boldsymbol{x}) \Pr(Y | \boldsymbol{x}).
    \label{eq:product_rule}
\end{equation}

While it is known as a rule-of-thumb that daily peaks tend to be observed under hot weather and in the afternoon, estimation of peak demand event frequency $\Pr(Y | \boldsymbol{x})$ is beyond the scope of this paper. 
In risk management, when $\Pr(Y | \boldsymbol{x})$ is unknown, a conservative practice is to adopt the worst-case assumption $\Pr(Y | \boldsymbol{x}) \equiv 1$, where peak demand events are assumed to always occur at every sub-interval.
Under this assumption, the predictions $\hat{\Pr}(y|Y, \boldsymbol{x})$ directly yields the joint distribution $\Pr(y, Y | \boldsymbol{x})$. 
While the identity \autoref{eq:product_rule} may resemble the structure of a Bayesian formulation, we emphasize that  Bayes' rule is not applied here to infer the probability of a cause given its effect.
Instead, the adoption of $\Pr(Y | \boldsymbol{x}) \equiv 1$ is a deliberate, risk-adverse modeling choice justified in the context of resource adequacy and reliability assessment.

\begin{figure}[htb!]
    \centering
    \includegraphics[width=\textwidth]{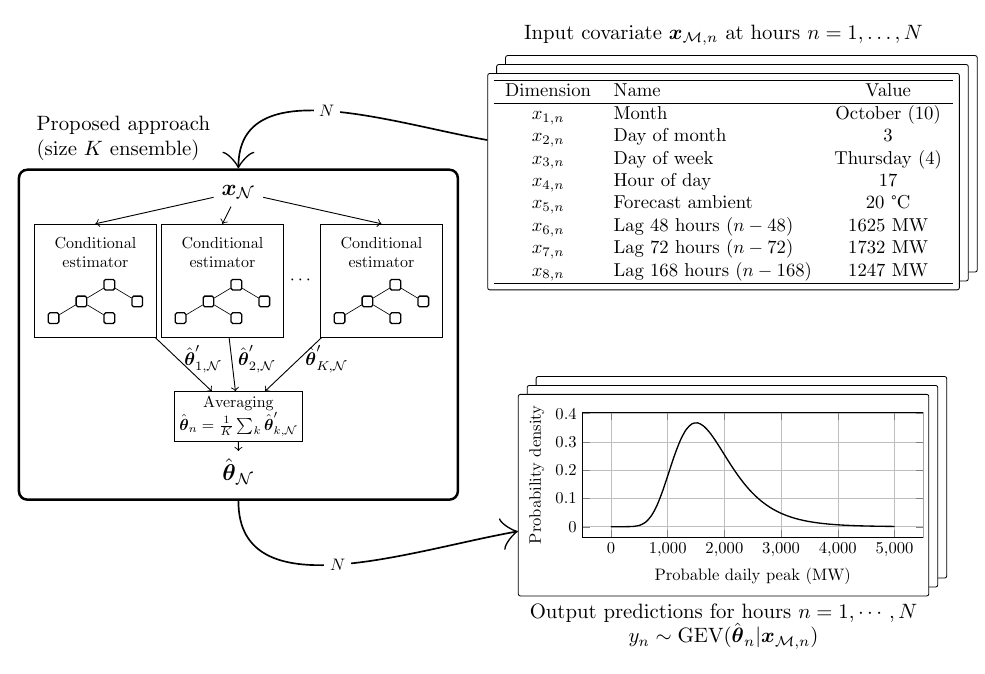}
    \caption{Example use case. The predictions are in the form of a conditional daily peak distribution $y \sim \text{GEV}(\hat{\boldsymbol{\theta}}|\boldsymbol{x})$ that is updated at every hour. The covariate elements and results in the artwork are examples intended to illustrate the mechanics of the proposed method and not to be taken as prescriptive implementation advice. GEV---generalized extreme value distribution.}
    \label{fig:model_overview}
\end{figure}

A typical application of the proposed approach is illustrated in \autoref{fig:model_overview}. The covariates exemplified in the artwork reflect common elements often chosen in time series forecasting, though the actual covariate structure is application-specific and informed by domain knowledge. 
Readers interested in a practical example are referred to the case study in \autoref{sec:case_study}, which demonstrates the covariate selection in a real-world setting. 
For the set of covariate dimensions $\mathcal{M}$, the set of hourly intervals $\mathcal{N}$, and the supplied covariate $\boldsymbol{x}_{\mathcal{M},\mathcal{N}}$ corresponding to hourly conditions, the proposed method produces estimates of the conditional daily peak distribution $\hat{\Pr}(y|Y, \boldsymbol{x}) = \text{GEV}(\hat{\boldsymbol{\theta}}|\boldsymbol{x})$ at every hour.

\subsection{Conditional framework for nonstationary extreme value densities}
For unknown initial distribution $D(\boldsymbol{x})$ that may change with covariate $\boldsymbol{x}$, sub-interval observations $z(\boldsymbol{x})$ under condition of $\boldsymbol{x}$ are assumed to be \gls{iid} realizations of the following stochastic process:
\begin{equation}
    z(\boldsymbol{x}) \sim D(\boldsymbol{x}) \label{eq:assumption}
\end{equation}

For the block maxima $ y(\boldsymbol{x}) =\max(z_1, \cdots, z_M|\boldsymbol{x})$ of $M$ \gls{iid} observations, according to the Fisher-Tippett-Gnedenko theorem, if as $M \rightarrow \infty$ the distribution of $y(\boldsymbol{x})$ is non-degenerate, it is well approximated by the \gls{GEV}, regardless of the initial distribution $D(\boldsymbol{x})$.
The \gls{DOA} refers to the set of initial distributions whose block maxima can be approximated by \gls{GEV}. For $p_1, \cdots, p_4\in(0,1)$, the necessary and sufficient condition for initial distribution $D$ with inverse \gls{CDF} $F^{-1}(z)$ to belong to the \gls{DOA} is \cite{castillo_extreme_2012}:
\begin{equation}
    \lim_{M\to\infty} \frac{F^{-1}(p_1^{1/M})-F^{-1}(p_2^{1/M})}{F^{-1}(p_3^{1/M})-F^{-1}(p_4^{1/M})} = 
    \frac{(-\log p_1)^{-\xi} - (-\log p_2)^{-\xi}}{(-\log p_3)^{-\xi} - (-\log p_4)^{-\xi}}
\end{equation}

The \gls{DOA} of \gls{GEV} encompasses a wide range of initial distribution families, making the statistical framework widely applicable. In general, block maxima of initial distributions with exponentially or sub-exponentially decaying upper tails, such as normal, exponential, $\log$-normal, Gamma, and Rayleigh, coverage to the type-I \gls{GEV} known as Gumbel, where $\xi=0$, those with polynomial decaying upper tails such as Cauchy, Pareto, and Student's $t$ converge to the type-II \gls{GEV} known as Fr\'echet, where $\xi>0$, and those with a finite upper bound, such as uniform or beta, converge to the type-III \gls{GEV} known as inverse Weibull, where $\xi<0$. 

For probable outcome $y$, location $\mu \in \mathbb{R}$, scale $\sigma > 0$, and shape $\xi \in \mathbb{R}$, the support of \gls{GEV} refers to $y$ that satisfies $1 + \xi \left( \frac{y-\mu}{\sigma} \right) > 0$. The \gls{PDF} of the \gls{GEV} is given by
\begin{equation}
\label{eq:gev_pdf}
    \Pr_\text{GEV}(y;\mu,\sigma,\xi) = \frac{1}{\sigma} \tau^{1+\xi} \exp(-\tau),
\end{equation}
where
\begin{equation}
\tau(y;\mu,\sigma,\xi) = 
\begin{cases}
    \left(1 + \xi \frac{y- \mu}{\sigma} \right)^{-\frac{1}{\xi}} &
    \xi \neq 0, \\
    \exp \left( - \frac{y - \mu}{\sigma} \right) & \xi = 0.
\end{cases}\label{eq:tau}
\end{equation}
When the block size $M$ is sufficiently large, the block maxima and minima can be modeled by
\begin{subequations}
\begin{align}
    \label{eq:max_follows_gev}
    y^\text{MAX}(\boldsymbol{x}) &\sim \text{GEV}(\mu^{\text{MAX}}, \sigma^{\text{MAX}}, \xi^{\text{MAX}}|\boldsymbol{x}), \\
    \label{eq:min_follows_gev}
    -y^\text{MIN}(\boldsymbol{x}) &\sim \text{GEV}(-\mu^{\text{MIN}}, \sigma^{\text{MIN}}, \xi^{\text{MIN}}|\boldsymbol{x}),
\end{align}\label{eq:max_min_follows_gev}
\end{subequations}
where
\begin{subequations}
\begin{align}
    y^\text{MAX}(\boldsymbol{x}) &= \max(z_1, \cdots, z_K|\boldsymbol{x}), \\
    y^\text{MIN}(\boldsymbol{x}) &= \min(z_1, \cdots, z_K|\boldsymbol{x}).
\end{align}
\end{subequations}

\subsection{Addressing autocorrelation}
\label{sec:autocorrelation}
When improperly modeled, autocorrelation in time series results in the weakening of the proposed framework due to violation of the \gls{iid} assumption \autoref{eq:assumption}. 
However, prior work shows that \gls{GEV} is still applicable under weak autocorrelation as long as the block maxima are effectively independent \cite{majumdar_extreme_2020}. 
Even with relatively strong autocorrelation with a slow rate of decay of the \gls{ACF}, the classical asymptotic behavior of \gls{EVA} still holds \cite{leadbetter_extremes_1983}.  
For example, \gls{GEV} is routinely applied in hydrology to estimate the probability of extreme floods from monthly rainfall maxima despite the month-to-month correlation.
Similar settings in power systems include \cite{ganger_statistical_2014}, where differencing reduces autocorrelation in ramp rates, and \cite{wang_high_2018}, where peaks are effectively independent. We argue that the conditional \gls{EVA} framework in \autoref{eq:max_min_follows_gev} is still applicable under strong autocorrelation, provided that one of the following configurations is used:
\begin{itemize}
\item Direct modeling of autocorrelation: For $\boldsymbol{z}_{<n}=[z_{n-r},\cdots, z_{n-2}, z_{n-1}]^\intercal$ representing the history of $z_n$ up to lag order $r$, assuming that strongly correlated lag orders are included in $\boldsymbol{z}_{<n}$, for exogenous factors $\boldsymbol\psi_n$, an autoregressive nonstationary process can be approximated as locally \gls{iid} using the unknown conditional distribution $z_n \sim D(\boldsymbol\psi_n,\boldsymbol{z}_{<n})$, which is in the form of \autoref{eq:assumption}. An example of this configuration is shown in \autoref{fig:model_overview}, where the predicted daily peak distribution is conditioned on strongly correlated historical demand.

\item Delegated modeling of autocorrelation: For $\hat{z}_n$ representing the point forecast produced by a base regression model known as the delegate, given the assumption that the residuals of such model are non-autocorrelated, for exogenous factors $\boldsymbol\psi_n$, an autoregressive nonstationary process can be approximated as locally \gls{iid} using the unknown conditional distribution $z_n \sim D(\boldsymbol\psi_n, \hat{z}_n)$, which is in the form of \autoref{eq:assumption}. This configuration is recommended when the delegate is already highly accurate, allowing the proposed approach to take advantage of the delegate when creating the peak demand probabilistic forecast. An example of this configuration is the PJM case study in \autoref{sec:case_study}.
\end{itemize}

It is demonstrated through the case study in \autoref{sec:case_study} that under worst-case baseline probability, this conditional \gls{EVA} framework gives rise to the exact amount of required day-ahead capacity to satisfy a given risk requirement and, compared to annual \gls{EVA} assessments, prevents over-commitment of generation resources and can potentially lead to a significant reduction of wholesale electricity prices.

    \section{Methodology}
\label{sec:methodology}
Given a set of peak conditions and historical peak demand in the form of covariate--target pairs $(\boldsymbol{x}_{\mathcal{M}\mathcal{N}}, y_{\mathcal{N}})$, the proposed approach aims to predict the parameters of the conditional distribution $\text{GEV}(\hat{\boldsymbol{\theta}}_{\mathcal{N}}|\boldsymbol{x}_{\mathcal{M},\mathcal{N}})$ by minimizing the disagreements between the predicted peak distribution and the actual peaks modeled by loss function $L$:
\begin{equation}
    \hat{\boldsymbol{\theta}}(\boldsymbol{x}) = 
    \argmin_{\boldsymbol{\theta}}
    \sum_{\forall (\boldsymbol{x}_{mn}, y_{n})}
    L(y_{n}; \boldsymbol{\theta})
\end{equation}

A well-known candidate for the loss function $L$ is the \gls{KL} \cite{kullback_information_1951}.
It quantifies the error of probabilistic forecast $\Pr_A(y;\hat{\boldsymbol \theta}_A)$ of distribution family $A$ against a reference probabilistic forecast $\Pr_B(y;\boldsymbol \theta_B)$ of distribution family $B$.
For observed outcome $y_n$, the ideal probabilistic forecast is a degenerate distribution that predicts $y_n$ with infinite probability density denoted by the Dirac delta function $\delta(\cdot)$. 
As shown in \autoref{eq:kl_divergence}, simplification of \gls{KL} against a degenerate distribution $B$ where $\Pr_B(y;\boldsymbol{\theta}_B)=\delta(y-y_n)$ leads to the log score, whose minimization results in the \gls{MLE}:
\begin{equation}
\label{eq:kl_divergence}
L=\text{KL}(\hat{A}, B) = 
\int \delta(y-y_n) \log \left(\frac{\delta(y-y_n)}{\Pr_A(y;\hat{\boldsymbol \theta}_A)} \right) 
= -\log \Pr_A(y_n;\hat{\boldsymbol \theta}_A)
\end{equation}

For \gls{GEV}, the above log score simplifies into
\begin{equation}
    L (y_n; \boldsymbol{\theta}) = \sigma - (1 + \xi) \log \tau + \tau.
    \label{eq:log_score}
\end{equation}

\begin{figure}[htb!]
    \centering
    \includegraphics[width=0.9\textwidth]{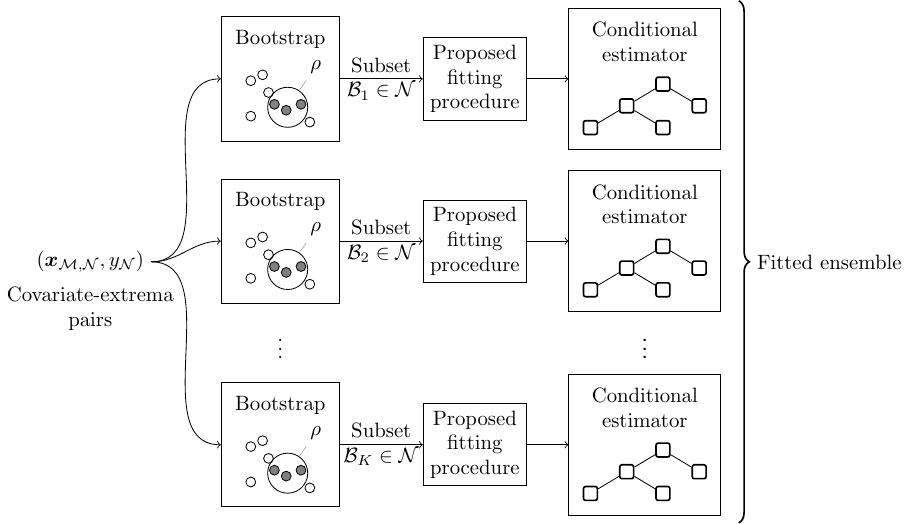}
    \caption{Ensemble fitting through bootstrapping, also known as resampling with replacement.}
    \label{fig:ensemble_fitting}
\end{figure}

As illustrated in \autoref{fig:ensemble_fitting}, the proposed approach adopts a bagging ensemble \cite{breiman_bagging_1996} of $K$ novel decision-tree-based estimators, each trained on subsets of the covariate--target pairs using a proposed procedure. 
The ratio $\rho \in (0, \infty)$ controls the size of the subsets relative to their parent.\footnote{Due to replacement, it is possible to create subsets larger than the original dataset.} 
During inference, which is exemplified in \autoref{fig:model_overview}, the member estimates are averaged to create the final estimate $\hat{\boldsymbol{\theta}}_\mathcal{N}(\boldsymbol{x}_{\mathcal{M}, \mathcal{N}})$:
\begin{equation}
    \hat{\boldsymbol{\theta}}_\mathcal{N}(\boldsymbol{x}_{\mathcal{M}, \mathcal{N}}) = \frac{1}{K} \sum_{k=1}^K \hat{\boldsymbol{\theta}}'_{k,\mathcal{N}} (\boldsymbol{x}_{\mathcal{M}, \mathcal{N}})
\end{equation}

\subsection{GEV parameter estimation}

The \gls{MLE} for \gls{GEV} does not have an explicit mathematical representation, and the procedure to compute it involves an iterative process starting from an initial estimate, guided by the gradient of the log score using a method now widely known as \gls{NGD} \cite{amari_natural_1998}. 
For \gls{GEV} Fisher information matrix $\boldsymbol{I}(\hat{\boldsymbol{\theta}})$ and (ordinary) gradient $\nabla L$ of the \gls{GEV} log score, the natural gradient $\tilde{\nabla} L$ is given by \cite{duan_ngboost_2020}:
\begin{equation}
    \tilde{\nabla} L(y_n;\hat{\boldsymbol{\theta}}) = 
    \boldsymbol{I}^{-1} (\hat{\boldsymbol{\theta}})
    \cdot
    \nabla L(y_n;\hat{\boldsymbol{\theta}}),
    \label{eq:natural_gradient}
\end{equation}
where $\tau$ is shown in \autoref{eq:tau}. 
For completeness, we reproduce the gradient components from \cite{dey_extreme_2016} and the Fisher information matrix from \cite{prescott_maximum_1980}.
The equations are not derived here but are provided for reference in \ref{sec:gev_gradient_formulas}.
It has been discovered that the \gls{MLE} for \gls{GEV} does not exist for $\xi \leq -1$ and does not possess asymptotic consistency when $-1 < \xi \leq -0.5$ \cite{dey_extreme_2016}. 
Furthermore, it is well-known that \gls{NGD} on \gls{GEV} is not numerically well-behaved---in addition to the need to test for singularity $\xi=0$, where \autoref{eq:ordinary_gradient} and \autoref{eq:fisher} must be replaced with their corresponding limits, due to poles in $\Gamma(x), \psi(x)$, numerical evaluation of $\boldsymbol{I}(\hat{\boldsymbol{\theta}})$ often results in an ill-conditioned matrix that is extremely difficult to invert. As a result, checking and reconditioning of exceptional numerical cases are required to prevent divergence, resulting in computational inefficiency. 

Due to the numerical issues inherent to the \gls{MLE}, the proposed approach instead performs parameter estimation through the \gls{PWM} method \cite{hosking_estimation_1985}. Despite having a small bias, the \gls{PWM} is found by \cite{martins_generalized_2000} to have a lower bias than \gls{MLE} at small sample sizes, and the advantage can be crucial under limited data availability. In fact, due to its superior stability, \gls{PWM} is often preferred over \gls{MLE} in fields such as hydrology  \cite{coles_introduction_2001}. We refer the reader to \cite{hosking_estimation_1985} for a comprehensive quantitative comparison between \gls{PWM} and \gls{MLE}. The \gls{PWM} estimate for \gls{GEV} is given by
\begin{subequations}\label{eq:pwm}
\begin{align}
\label{eq:pwm_mu}
\hat{\xi} &= -7.8590c - 2.9554c^2, \\
\label{eq:pwm_sigma}
\hat{\sigma} &= 
\frac{(b_0 - 2b_1)\hat{\xi}}{\Gamma(1 - \hat{\xi}) (1 - 2^{\hat{\xi}})}, \\
\label{eq:pwm_xi}
\hat{\mu} &= 
b_0 - \frac{\hat{\sigma}}{\hat{\xi}} (\Gamma(1 - \hat{\xi}) - 1),
\end{align}
\end{subequations}
and for $y'_1,\cdots, y'_N$ representing the block maxima sorted by value in ascending order, the quantities $b_0, b_1, b_2, c$ are given by
\begin{subequations}
\begin{align}
\label{eq:pwm_b0}
b_0 &= \frac{1}{N} \sum_{j=1}^{N} y'_j, \\
\label{eq:pwm_b1}
b_1 &= \frac{1}{N} \sum_{j=1}^{N} \frac{j-1}{N-1} y'_j, \\
\label{eq:pwm_b2}
b_2 &= \frac{1}{N} \sum_{j=1}^{N} \frac{(j-1)(j-2)}{(N-1)(N-2)} y'_j, \\
\label{eq:pwm_c}
c &= \frac{2b_1 - b_0}{3b_2 - b_0} - \frac{\log 2}{\log 3}.
\end{align}
\end{subequations}

\subsection{Construction of the conditional estimators}
\label{sec:conditional-estimators}
We adopted the procedure in \cite{breiman_classification_2017}, known as \gls{CART}. It is a recursive partitioning algorithm that, during fitting, dichotomizes the covariate-target pairs using a set of automatically discovered splitting rules, which are organized in the form of a tree. In the prediction phase, the splitting rules are applied to the given covariate in topological order, resulting in the activation of one of the leaf partitions whose estimate gives rise to the prediction. 
For a set of covariate dimensions $\mathcal{M}$, parent observation indices $\mathcal{P}$ and parent partition $(\boldsymbol{x}_{\mathcal{M},\mathcal{P}}, y_\mathcal{P})$, the child observation indices $L, \mathcal{R}$ resulting from splitting the parent in covariate dimension $m$ on threshold $u$ are:
\begin{subequations}\label{eq:children}
\begin{align}
    \mathcal{L} &=\{p|p \in \mathcal{P} \land \boldsymbol{x}_{m,p} \leq u \} \label{eq:left_child} \\
    \mathcal{R} &=\{p|p \in \mathcal{P} \land \boldsymbol{x}_{m,p} > u \} \label{eq:right_child}
\end{align}
\end{subequations}

The tuple $(m, u)$ is known as the splitting rule of the partition. In \gls{CART}, the selection of the optimal splitting rule is based on maximizing the reduction in squared error impurity, resulting in children in which target observations are similar in value. However, targets in the same partition do not necessarily have the same distribution. We hence modified \gls{CART}, using the \gls{GEV} log score \autoref{eq:log_score} to select the optimal splitting rule. 
For covariate-target pairs $(\boldsymbol{x}_{\mathcal{M}, \mathcal{N}}, y_\mathcal{N})$, the procedure to construct the conditional estimators is described as follows:
\begin{enumerate}[S1]
\item\label{item:tree_fitting1} The modified \gls{CART} begins with the creation of a root node to which all covariate--target pairs are assigned.
\item\label{item:tree_fitting2}
A covariate dimension is selected, starting with $m=1$. The unique values in the dimension are sorted, considering the midpoints as candidates for the splitting threshold. For each candidate splitting rule $(m, u)$, transient partitions $y_L, y_\mathcal{R}$ are created via \autoref{eq:children}. The \gls{GEV} \gls{PWM} estimates and log scores are computed on $y_L, y_\mathcal{R}$ via \autoref{eq:pwm} and \autoref{eq:log_score}, respectively. For parent log score $L_p$, and children log scores $L_l, L_r$, we define the impurity drop $T$ as the reduction of the children log scores relative to the parent:
\begin{equation}
T = \frac{1}{L_p}(L_p - L_r - L_l)
\end{equation}
The splitting rule that results in the maximum impurity drop is selected as the optimal rule of the partition.
\item\label{item:tree_fitting3} The leaf nodes are scanned, and the leaf whose optimal rule results in the maximum impurity reduction is selected as the growing leaf. The optimal splitting rule of the leaf is finally applied to create the child partitions, which are inserted under the leaf. Step \ref{item:tree_fitting2} is repeated recursively on the new children.
\end{enumerate}

While \gls{CART} adopts post-processing procedures known as pruning to trim certain branches to reduce the complexity of the tree, we adopt early termination, also known as pre-pruning, to suppress over-fitted branches during tree construction. The conditions for early termination are:
\begin{itemize}
\item The optimal threshold search in Step \ref{item:tree_fitting2} is limited to those that result in partitions of at least a specified minimum size.
\item No further children are allowed under a branch if the maximum impurity reduction of its leaf is lower than a specified limit $T_\text{crit}$. 
\item The number of growing iterations is limited to a specified maximum.
\end{itemize}

\begin{figure}
    \centering
    \includegraphics[width=0.9\linewidth]{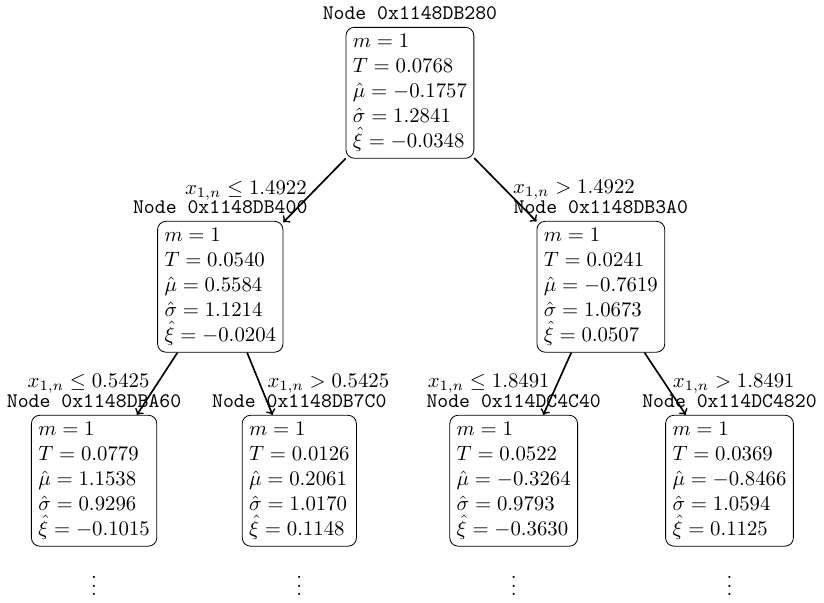}
    \caption{Structure of one of the conditional estimators created during fitting. The artwork is intended to illustrate the fitting procedure. Deeper nodes are not shown to conserve ink.}
    \label{fig:example_tree}
\end{figure}

To illustrate the fitting procedure, the runtime data from one of the conditional estimators during fitting of the example in \autoref{sec:example} is shown in \autoref{fig:example_tree}. The artwork illustrates only the top few nodes of only one of the conditional estimators, while the full ensemble consists of many such trees.
Readers are encouraged to cross-examine \autoref{fig:example_tree} with the procedure described in Steps \ref{item:tree_fitting1} to \ref{item:tree_fitting3} to understand the fitting process. 
It can be observed from \autoref{fig:example_tree} that as the tree grows deeper, each split conditions the block maxima on increasingly specific subsets of the covariate. Consequently, the fitted GEV parameters evolve from broad, population-level estimates near the root toward specialized distributions in the leaves that capture localized effects. This hierarchical refinement allows the parameter estimates to specialize progressively, with deeper nodes emphasizing finer variations in extreme densities.

\paragraph{Summary of hyperparameters}
The complete list of hyperparameters and their selection guidelines are shown in \autoref{tab:hyperparameters}. 
Typical of machine learning methods, no closed-form solutions exist for optimal values, and hyperparameter selection is largely empirical and application-dependent, based on trial and error.

\begin{table}[htb!]
    \centering
    \scalebox{0.8}{
    \renewcommand{\arraystretch}{1.7}
    \begin{tabularx}{1.2\textwidth}{cXc}
        \hline
        Hyperparameter &
        Purpose and
        Recommendation &
        Typical Value
        \\
        \hline
        \texttt{ensemble\_size} &
        Number of trees in the bagging ensemble. Affects estimate stability. Increase if the estimates on the same training set vary greatly between training invocations. &
        $50$
        \\
        $\rho$ &
        Resampling ratio. Controls the size of the training subsets relative to the training set. Start with 1 and decrease to reduce training time on multi-decade training sets. Set to larger than 1 to improve estimate stability in conjunction with increasing \texttt{ensemble\_size}. &
        $1$
        \\
        \texttt{max\_n\_splits} &
        Maximum number of growing iterations for each tree. Limits the tree complexity. Set to as large as computationally feasible. & 
        $40$
        \\
        \texttt{min\_partition\_size} &
        Minimum number of observations assigned to leaf partitions. A larger value provides more accurate estimates and prevents overfitting, but causes underfitting in smaller datasets. &
        $\ge 20$
        \\
        $T_\text{crit}$ &
        Minimum impurity reduction. Suppresses creation of statistically insignificant or overfitted branches. A large value prevents overfitting but causes underfitting. Start with a large value and decrease as needed. &
        $0.0001-0.01$
        \\
        \hline
    \end{tabularx}
    }
    \caption{Hyperparameters, purpose, practical recommendations, and typical values.}
    \label{tab:hyperparameters}
\end{table}
    \section{Application-neutral example}
\label{sec:example}
To illustrate the general applicability of the proposed approach, we first present an application-neutral example before turning to the power system case study in the next section. This example is not a forecasting task but a controlled training experiment designed solely to assess the quality of fit. It demonstrates the model’s ability to approximate highly nonlinear variations in the \gls{GEV} parameters under conditions where the true reference distribution is known. Because the ground truth is available, model performance can be evaluated directly against it, making a validation or test split unnecessary. This synthetic setup, therefore, enables rigorous comparison of the proposed method with benchmark estimators under well-defined conditions.
The data for the example consists of $\mathcal{N}=\{1,\cdots, 1000\}$ covariate--target pairs where the covariate $\boldsymbol{x}_{\mathcal{M},\mathcal{N}}$ has one dimension $\mathcal{M}=\{1\}$ and the pairs are synthesized according to the following rules:
\begin{subequations}
\begin{align}
    x_{1,\mathcal{N}} &\in [0, \pi], \\
    y_\mathcal{N} &\sim \text{GEV}(\mu_\mathcal{N}, \sigma_\mathcal{N}, \xi_\mathcal{N} | \boldsymbol{x}_{\mathcal{M},\mathcal{N}}),
\end{align}
\end{subequations}
where the instantaneous distribution parameters are given by
\begin{subequations}\label{eq:synthetic_example}
\begin{align}
    \mu_n(x_{1,n}) &= \cos(1.23 x_{1,n}) + 0.3 \cos(4.56 x_{1,n}),\\
    \xi_n(x_{1,n}) &= 0.3 \cos(5.67 x_{1,n}), \\
    \sigma_n(x_{1,n}) &= 1 + 0.1 \cos(6.78 x_{1,n}).
\end{align}
\end{subequations}

\begin{figure}[htb!]
    \centering
    \includegraphics[width=\textwidth]{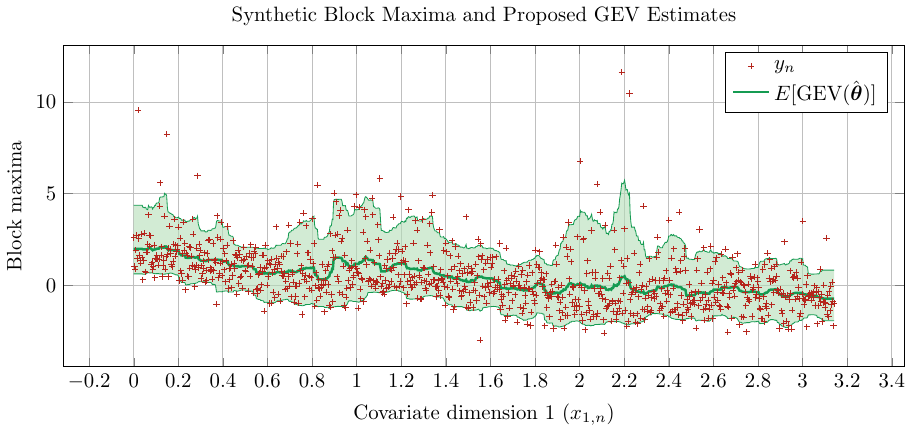}
    \caption{The synthesized block maxima and the nonstationary GEV estimated using the proposed approach visualized in the form of its expectation and the $90\%$ prediction interval (shaded region). GEV---generalized extreme value distribution.}
    \label{fig:synthetic_observations}
\end{figure}

\begin{figure}[htb!]
    \centering
    \includegraphics[width=\textwidth]{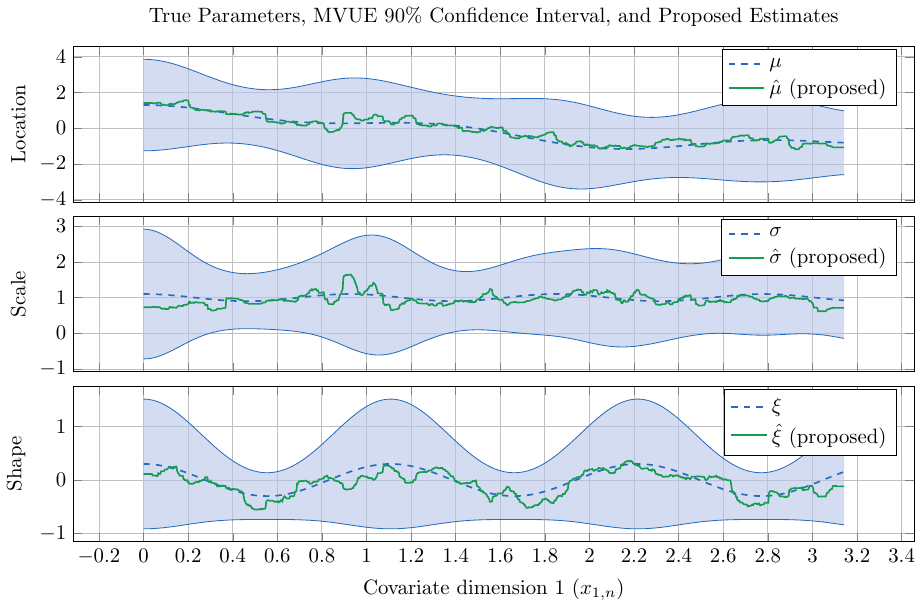}
    \caption{The true parameters of the distribution that synthesized the block maxima, the $90\%$ confidence interval of a theoretical and ideal MVUE (shaded region), and the estimates produced by the proposed approach. MVUE---minimum variance unbiased estimator.}
    \label{fig:synthetic_params}
\end{figure}

The hyperparameters are selected following the guidelines in \autoref{tab:hyperparameters}, and the selected values are shown in \autoref{tab:example-hyperparameters}. On a typical modern personal computer, fitting the example takes approximately 53 seconds.
The \gls{GEV} distribution estimated using the proposed approach is visualized in the form of the $90\%$ prediction interval against the synthesized block maxima in \autoref{fig:synthetic_observations}. 
The parameter estimates are shown in \autoref{fig:synthetic_params}. 

\begin{table}[htb!]
    \centering
    \scalebox{0.8}{
    \begin{tabular}{lr}
        \hline
        Hyperparameter &
        Selected Value
        \\
        \hline
        \texttt{ensemble\_size} &
        $50$
        \\
        $\rho$ &
        $1$
        \\
        \texttt{max\_n\_splits} &
        $40$
        \\
        \texttt{min\_partition\_size} &
        $20$
        \\
        $T_\text{crit}$ &
        $0.0001$
        \\
        \hline
    \end{tabular}
    }
    \caption{Value of hyperparameters in the application-neutral example.}
    \label{tab:example-hyperparameters}
\end{table}

A robust estimator should produce consistent estimates on samples drawn from the same distribution, despite the observations being different each time. The bias relates to the asymptotic accuracy of the estimator, and the variance relates to the stability of the estimate. The achievable minimum asymptotic variance of the estimates produced by an unbiased estimator known as the \gls{MVUE} is given by the \gls{CRB} as follows:
\begin{equation}
\text{Var}(\hat{\boldsymbol{\theta}}) \geq I^{-1}(\boldsymbol{\theta}),
\label{eq:crb}
\end{equation}
where $I(\boldsymbol{\theta})$ is the \gls{GEV} Fisher information given by \autoref{eq:fisher}. An unbiased estimator that achieves the \gls{CRB} is considered statistically most efficient. To contrast the proposed approach, we graphically represent the $90\%$ parameter confidence intervals of the \gls{MVUE} calculated using the diagonal elements of \autoref{eq:crb} in the shaded regions of \autoref{fig:synthetic_params}. Since the \gls{PWM} method is biased, a test for significant parameter deviation, such as the Wald test \cite{wald_tests_1943}, would not be well-calibrated. Nonetheless, it can be observed that estimates of the proposed approach fall within the interval, a graphical indication that such tests would likely be unable to reject the hypothesis that there are no statistically significant deviations between the proposed method and the true parameters.

\subsection{Comparison with a state-of-the-art parametric competitor}
\Gls{GAMLSS} \cite{rigby_generalized_2005} provides a flexible framework that can approximate the parameters of a variety of distribution families as smooth functions of the covariate. It is widely used in applications such as climatology \cite{dabernig_simultaneous_2017} and public health \cite{giacomet_distributional_2023} to model the variation of probability densities in response to environmental factors. However, GAMLSS comes with several notable practical drawbacks:

\begin{itemize}
    \item \Gls{GAMLSS} requires the labor-intensive manual specification of formulas for all three \gls{GEV} parameters through trial-and-error, which makes the forecasting model prone to misspecification.
    
    \item Spline-based function approximations are not well-suited to capturing discontinuous or highly nonlinear relationships that may be encountered during peak demand probabilistic modeling.

    \item \Gls{GAMLSS} can suffer from over-smoothing or under-fitting, where the estimates fail to adequately capture the behavior of the distribution parameters.
\end{itemize}

To empirically validate these limitations, the proposed method is compared with \gls{GAMLSS} \cite{rigby_generalized_2005} against the true parameters.
All three \gls{GEV} parameters in the \gls{GAMLSS} model are approximated using penalized B-splines of the covariate $\boldsymbol{x}_{1,n}$. 
As observed, although the improvements in location, scale, and shape parameter estimates may appear visually modest, \gls{GAMLSS} estimates tend to underfit, especially where covariate effects are strong. This is especially obvious for the $\sigma$ and $\xi$ parameters. The proposed method more closely follows the true parameter curves. 
While not universally superior in every region, the proposed method usually provides better alignment, especially in the $\sigma$ and $\xi$ estimates.

\begin{figure}[htb!]
    \centering
    \includegraphics[width=\textwidth]{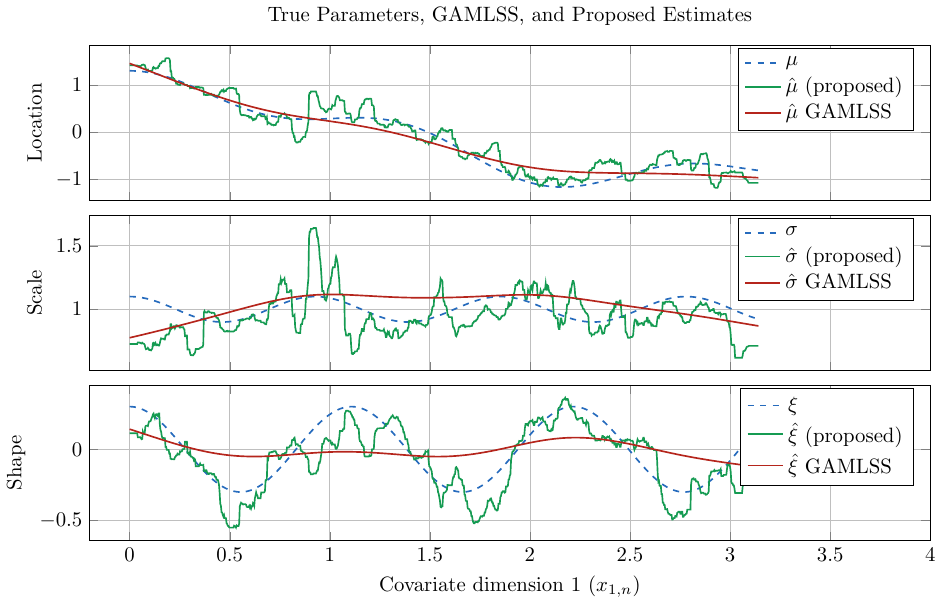}
    \caption{The true parameters of the distribution that synthesized the block maxima, the estimates produced by the proposed approach, and the estimates produced by GAMLSS. GAMLSS---generalized additive model for location, scale, and shape.}
    \label{fig:synthetic_params_gamlss}
\end{figure}

The \gls{CRPS} is adopted for a more precise quantitative comparison. The \gls{CRPS} is a robust scoring metric often used in forecast verification in numerical weather prediction. It compares the \gls{CDF} of the predicted density against an ideal \gls{CDF} corresponding to each observation. 
For gamma function $\Gamma(x) = \int_0^\infty \tau^{x-1} \exp (-\tau) d\tau$, lower incomplete gamma function $\Gamma_l(s,x)=\int_0^x t^{s-1} \exp(-t) dt$, exponential integral function $\text{Ei}(x)=\int_{-\infty}^x e^t/t dt$,  and Euler--Mascheroni constant $\gamma_\text{EM} = -\psi(1) \approx 0.5772$, the \gls{CRPS} formula for \gls{GEV} is derived by \cite{friederichs_forecast_2012} as follows:
\begin{equation}
    \text{S}(y, \hat{\boldsymbol{\theta}}) = 
    \begin{cases}
        \begin{aligned}
            &\left(\hat{\mu} - y - \frac{\hat{\sigma}}{\hat{\xi}} \right) 
            \left(1 - 2 F_{\xi\neq 0}(y) \right) \\
            &\quad -\frac{\hat\sigma}{\hat\xi} \left( 
            2^{\hat\xi} \Gamma(1 - \hat\xi) - 2 \Gamma_l (1 - \hat\xi, -\log F_{\xi\neq 0}(y)) \right),
        \end{aligned} & \hat\xi \neq 0, \\
        \hat{\mu} - y + \hat{\sigma} (\gamma_\text{EM} - \log 2) - 2\hat{\sigma} \text{Ei} (\log F_{\xi=0}(y)), & \hat\xi = 0,
    \end{cases}
    \label{eq:crps_gev}
\end{equation}
where $F(y)$ is the \gls{CDF} of \gls{GEV}, given by
\begin{equation}
    F(y) = \exp(-\tau), 
\end{equation}
and $\tau$ is shown in \autoref{eq:tau}. The \gls{CRPS} of the proposed method and \gls{GAMLSS} on the previously described dataset is listed in \autoref{tab:ours_and_gamlss_crps}. The scores shown are the average \gls{CRPS} over all observations, and the per-observation \gls{CRPS} is calculated via \autoref{eq:crps_gev}. A lower score indicates better agreement between the observations and the \gls{CDF} of the estimated density.

\begin{table}[htb!]
    \centering
    \begin{tabular}{cc}
        \hline 
        GAMLSS CRPS & \textbf{Proposed CRPS} \\
        \hline
        0.719 & 0.677 \\
        \hline
    \end{tabular}
    \caption{Comparison of the CRPS between the proposed approach and GAMLSS in the application-neutral example. CRPS---continuous ranked probability score, GAMLSS---generalized additive model for location, scale, and shape.}
    \label{tab:ours_and_gamlss_crps}
\end{table}

\subsection{Exploring the limitation of a quantile-based non-parametric competitor}
\begin{figure}[htb!]
\centering
\begin{minipage}[t]{0.48\textwidth}
    \centering
    \includegraphics[width=\textwidth, page=1]{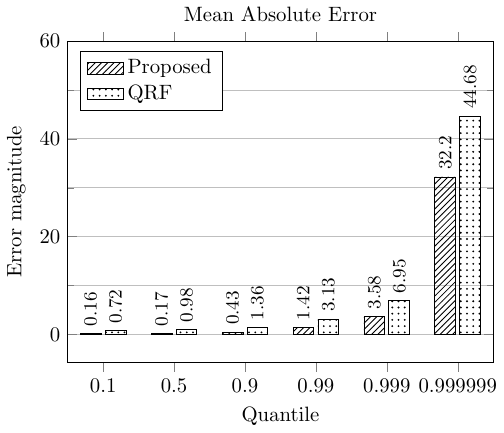}
\end{minipage}
\begin{minipage}[t]{0.48\textwidth}
    \centering
    \includegraphics[width=\textwidth, page=2]{999-fig-009.pdf}
\end{minipage}
\caption{The errors between quantile estimates given by QRF and the proposed approach against the true quantile target of the distribution that synthesized the block maxima. QRF---quantile regression forest.}
\label{fig:synthetic_example_quantile_losses}
\end{figure}

\begin{figure}[htb!]
\centering
\begin{minipage}[t]{\textwidth}
    \centering
    \includegraphics[width=\textwidth, page=1]{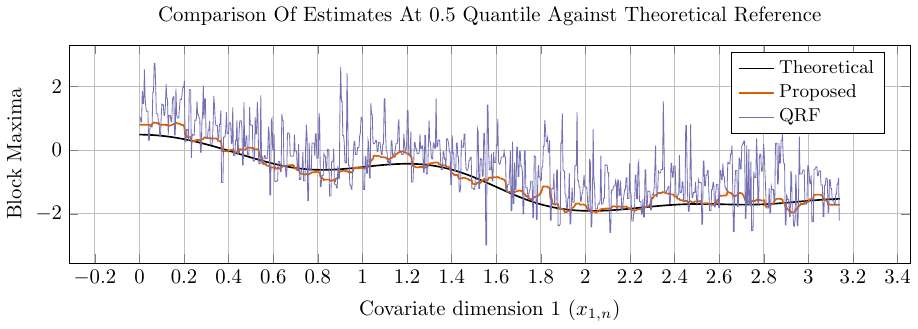}
\end{minipage}
\begin{minipage}[t]{\textwidth}
    \centering
    \includegraphics[width=\textwidth, page=2]{999-fig-010.pdf}
\end{minipage}
\begin{minipage}[t]{\textwidth}
    \centering
    \includegraphics[width=\textwidth, page=3]{999-fig-010.pdf}
\end{minipage}
\begin{minipage}[t]{\textwidth}
    \centering
    \includegraphics[width=\textwidth, page=4]{999-fig-010.pdf}
\end{minipage}
\caption{The quantile estimates given by QRF and the proposed approach, and the true quantile target of the distribution that synthesized the block maxima. Note that the vertical axis differs in scale considerably between plots. QRF---quantile regression forest.}
\label{fig:synthetic_black_swan}
\end{figure}

Since the block maxima in the example follow a nonlinear relationship with the covariate $\boldsymbol{x}_{\mathcal{M},\mathcal{N}}$, linear approaches such as \gls{QR} would discernibly be out-competed by the proposed approach, which can model nonlinear relationships. 
To present a fair case-by-case comparison across different quantiles, quantile estimates of the block maxima from the proposed approach is computed using using the inverse \gls{CDF} shown in \autoref{eq:inverse_cdf} and compared them with the true quantile targets based on \autoref{eq:synthetic_example} and quantile estimates of \gls{QRF} \cite{johnson_quantile-forest_2024}. The results are shown in \autoref{fig:synthetic_example_quantile_losses} and \autoref{fig:synthetic_black_swan}.
\begin{equation}
F^{-1}(\alpha; \boldsymbol{\theta})=
\begin{cases}
    \mu - \sigma \log(-\log \alpha) & \xi = 0, \\
    \mu + \frac{\sigma}{\xi} \left((-\log \alpha)^{-\xi} -1 \right) & \xi \neq 0.
\end{cases}
\label{eq:inverse_cdf}
\end{equation}

It can be observed in \autoref{fig:synthetic_example_quantile_losses} that the proposed approach outperforms \gls{QRF} in all evaluated quantiles and especially at the \gls{LPHC} quantiles. In \autoref{fig:synthetic_black_swan}, for quantiles $0.1, 0.5$, and $0.9$, the estimates of \gls{QRF} are especially noisy compared to estimates of the proposed approach and the true block maxima targets, suggesting some level of over-fitting by \gls{QRF}. 
Given that an over-fitted \gls{QRF} possesses increased sensitivity to extreme observations, for \gls{LPHC} quantiles $0.999$ and $0.999999$, however, the \gls{QRF} estimates deviate significantly from the true targets, while the quantile estimates produced by the proposed approach are in significantly better agreement with the true targets.
The scarcity of extreme observations limits the ability of quantile-based methods, including highly sophisticated deep-learning-based quantile regression methods, to accurately approximate distribution tails. 
Quantile methods are often out-competed in terms of accuracy and trustworthiness by \gls{EVA} in critical applications such as finance and insurance underwriting. 
For more comprehensive comparisons between quantile-based methods and \gls{EVA}, we refer the reader to \cite{mcneil_quantitative_2015}. 

    \section{Case study---day-ahead scheduling capacity determination}
\label{sec:case_study}
\begin{figure}[htb!]
    \centering
    \includegraphics[width=0.7\textwidth]{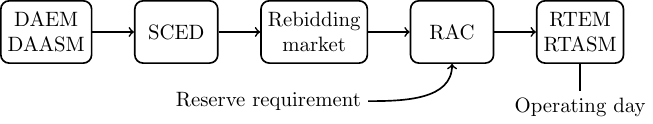}
    \caption{Simplified block diagram of the scheduling workflow of PJM. DAEM---day-ahead energy market, DAASM---day-ahead ancillary service market, SCED---security-constrained economic dispatch, RAC---reliability assessment and commitment, RTEM---real-time energy market, RTASM---real-time ancillary service market.}
    \label{fig:pjm_market}
\end{figure}

The PJM interconnection is a \gls{RTO} serving 65 million people across 13 states and the District of Columbia. The system demand ranges between $\SI{80}{\giga\watt}$ to $\SI{150}{\giga\watt}$, the summer peak in 2024 exceeding $\SI{152}{\giga\watt}$. This section presents a case study of the proposed approach applied to the PJM Interconnection. Although the case study focuses on PJM, the proposed framework is model-agnostic and system-independent, relying solely on historical peak demand data and exogenous covariates. Consequently, its application to other power systems is straightforward, requiring only minor methodological adjustments and metric-specific adaptations to align with the reliability indicators used.
In this case study, we demonstrate day ahead capacity determination based on the \gls{LOLP}.
 The simplified block diagram of the scheduling workflow of PJM is shown in \autoref{fig:pjm_market}. 

\begin{figure}[htb!]
    \centering
    \includegraphics[width=0.55\textwidth]{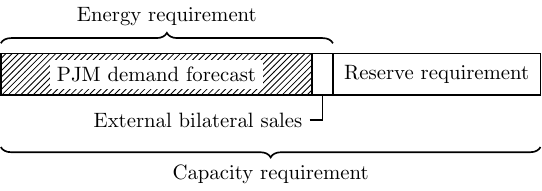}
    \caption{Composition of the day-ahead scheduling capacity.}
    \label{fig:pjm_capacity}
\end{figure}

According to PJM Manual 11 \cite{pjm_interconnection_pjm_2021}, day-ahead bids close at 11:00, followed by market clearing through the \gls{SCED} engine, which posts generation schedules, \gls{LMP}, and \gls{MCP} by 13:30.
Participants may rebid until 14:45, when PJM conducts the \gls{RAC} to ensure sufficient reserves for the next day.
To prepare for contingencies and maintain energy balance, these reserves, shown in \autoref{fig:pjm_capacity}, are scheduled and placed on standby during operations. 
Because the \gls{RAC} is the final opportunity to commit capacity, the proposed forecasts must align with the 14:45 timeline to be operationally relevant. 
In this case study, the method is applied to determine risk-based \gls{DASC} requirements, focusing on total system demand within the PJM \gls{RTO} region. For simplicity, external bilateral sales and scheduled interchanges with neighboring balancing authority areas are excluded.

\subsection{Data collection and train--test split}
The PJM Data Miner 2 platform \cite{pjm_interconnection_pjm_2025} provides publicly accessible datasets, from which demand \gls{HIA}, hourly demand forecasts, and historical day-ahead temperature forecasts for study years 2020--2024 were obtained. 
Since the platform restricts downloads to a five-year window, creating a separate validation set would substantially reduce the training sample. 
To balance data availability with evaluation rigor, training was performed using data from 2020--2023 without a separate validation set, and model performance was assessed on a hold-out test set covering the year 2024.

\subsection{Fitting and fit quality}

\begin{figure}[htb!]
    \centering
    \includegraphics[width=0.7\textwidth, page=1]{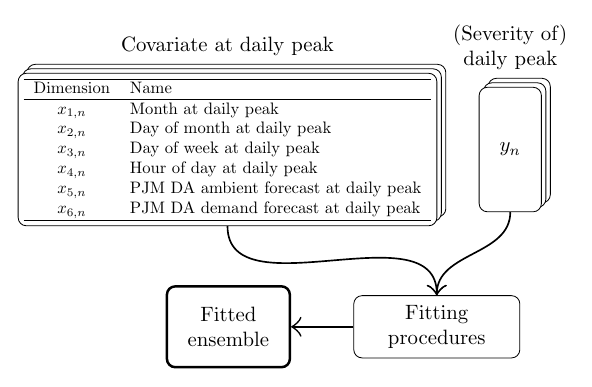}
    \caption{The fitting data contains only the daily peaks and the corresponding covariate, whose structure is shown in the figure. DA---day-ahead.}
    \label{fig:pjm_training}
\end{figure}

\paragraph{Covariate structure}
As illustrated in \autoref{fig:pjm_training}, daily peak demand is modeled as $y \sim \text{GEV}(\boldsymbol{\theta}|\boldsymbol{x})$, where the covariates $\boldsymbol{x}$ consist of calendar variables, day-ahead temperature, and PJM base demand forecasts. 
The selection of covariates is the result of widely accepted general practices consistent with prior load forecasting studies.
Calendar variables in the covariate capture recurrent temporal effects such as seasonality and weekly cycles on the peak demand, and ambient temperature reflects weather-driven demand sensitivity.
The PJM base demand forecasts, updated every six hours, provide single-valued hourly predictions for the following day. These base forecasts, generated using neural networks and pattern-matching methods \cite{pjm_interconnection_pjm_2021}, are highly accurate. 
Based on the recommendation in \autoref{sec:autocorrelation}, delegated covariate modeling can take advantage of the highly accurate PJM base forecasts when creating the peak demand probabilistic forecasts by embedding the base forecasts in the covariate.
To align with the PJM market process in \autoref{fig:pjm_market}, the capacity requirements must be available at least 36 hours in advance for the \gls{RAC} run at 14:45. Therefore, the PJM base forecasts issued at 05:45 ahead of the operating day were adopted.

The hyperparameters are selected using the guidelines in \autoref{tab:hyperparameters}, and the selected values are shown in \autoref{tab:pjm-hyperparameters}. 
On a typical modern personal computer, fitting the case study takes approximately 67 seconds.

\begin{table}[htb!]
    \centering
    \scalebox{0.8}{
    \begin{tabular}{lr}
        \hline
        Hyperparameter &
        Selected Value
        \\
        \hline
        \texttt{ensemble\_size} &
        $50$
        \\
        $\rho$ &
        $1$
        \\
        \texttt{max\_n\_splits} &
        $40$
        \\
        \texttt{min\_partition\_size} &
        $30$
        \\
        $T_\text{crit}$ &
        $0.05$
        \\
        \hline
    \end{tabular}
    }
    \caption{Value of hyperparameters in the PJM case study.}
    \label{tab:pjm-hyperparameters}
\end{table}

\paragraph{Residual analysis}
The fitting residual $r_\mathcal{N}$ is the difference between the statistical expectation of the estimated peak distribution and the actual daily peak $y_\mathcal{N}$ in the training dataset:
\begin{equation}
    r_\mathcal{N} \triangleq \hat{y}_\mathcal{N} - y_\mathcal{N}
    = 
    \mathbb{E}\left[\text{GEV}(\hat{\boldsymbol{\theta}}|\boldsymbol{x}_{\mathcal{M}, \mathcal{N}})\right] - y_\mathcal{N},
\end{equation}
and the expectation is given by
\begin{equation}
    \mathbb{E}\left[\text{GEV}(\boldsymbol{\theta})\right] = 
    \begin{cases}
        \mu + \frac{\sigma}{\xi} \left( \Gamma(1-\xi) - 1\right) & \xi \in (-\infty, 1)\setminus \{0\},\\
        \mu + \sigma \gamma_\text{EM} & \xi = 0, \\
        \infty & \xi \geq 1.
    \end{cases}
\end{equation}

\begin{figure}[ht!]
    \centering
    \begin{minipage}[t]{0.49\textwidth}
        \centering
        \includegraphics[width=\textwidth, page=1]{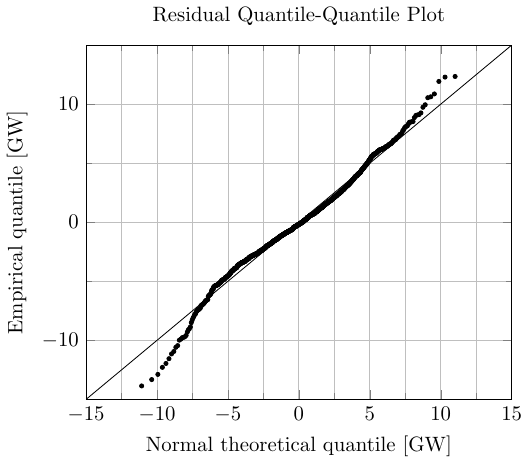}
    \end{minipage}
    \hfill
    \begin{minipage}[t]{0.47\textwidth}
        \centering
        \includegraphics[width=\textwidth, page=2]{999-fig-014.pdf}
    \end{minipage}
    \caption{Quantile-quantile plot and histogram of the daily peak residuals in study years 2020--2023.}
    \label{fig:pjm_residual}
\end{figure}

The residuals of a robust forecasting model should follow a stationary normal distribution with a small mean and variance. It can be observed from \autoref{fig:pjm_residual} that the residuals in study years 2020--2023 appear to follow a normal distribution, indicating that the model fitting has been performed correctly, the selection of covariates appears to be sufficient, and that the choice of hyperparameters appears to be appropriate to capture the underlying dynamics without underfitting.

\paragraph{General test performance}
After fitting, covariates of test year 2024 are supplied to the proposed model, resulting in forecasts for the entire study year.
The forecasts are in the form of \gls{GEV} parameter estimates for the daily peak density, and the predicted density is updated hourly as conditions change.
Levaging the highly accurate PJM point forecast in the covariate, the \gls{MAE} of the expectation of the forecasted daily peak density against the actual daily peaks in test year 2024 is $\SI{2814.28}{\mega\watt}$, or  $1.84\%$ of the $\SI{153}{\giga\watt}$ total system demand. The mean under-forecasted forecasting error of the daily peak is $\SI{1850.15}{\mega\watt}$, or $1.21\%$ of the total system demand.
These results indicate that the fitted ensemble predicts the dynamics of peak demand density with high fidelity while maintaining low bias relative to system scale.

\subsection{Results: day-ahead scheduling capacity}

\begin{figure}[htb!]
    \centering
    \includegraphics[width=0.7\textwidth]{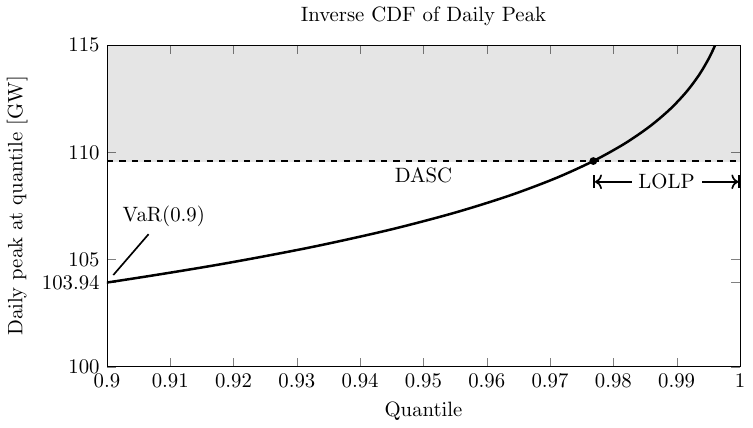}
    \caption{Relationship between the inverse CDF, the VaR, the DASC, and the LOLP. CDF---cumulative distribution function, VaR---value-at-risk, DASC---day-ahead scheduling capacity, LOLP---loss-of-load probability.}
    \label{fig:quantile_function}
\end{figure}

According to PJM Manual 13 \cite{pjm_interconnection_manual_2025}, the 30-minute reserve requirement is determined from the annual peak load forecast, adjusted for the under-forecasted load-forecasting error and generator \gls{FOR}. However, reserve planning at the reliability standard level is governed by the broader ``one day in ten years'' guideline specified by \gls{NERC} standard BAL-502-RF-03 \cite{north_american_electric_reliability_corporation_bal-502-rf-03_2025} based on the \gls{LOLP}.
The \gls{NERC} guideline requires the annual \gls{LOLP} to be equal to 0.1 and generalizes into daily \gls{LOLP} risk requirement $\eta$ as follows:
\begin{equation}
    \eta = \frac{\SI{0.1}{\si{Events}\per\si{year}}}{\SI{365}{\si{days}\per\si{year}}} 
    = \SI{273.9e-6}{\si{Events}\per\day}
    \label{eq:nerc_risk}
\end{equation}
The rule \autoref{eq:nerc_risk} can be used to calculate the required \gls{DASC}.
For the event $Y$ that the daily peak occurs at hourly condition $\boldsymbol{x}_n$, joint probability $\Pr(y,Y| \boldsymbol{x}_n)$ of daily peak of severity $y$ occurring under hourly condition $\boldsymbol{x}_n$, the \gls{DASC} that can accommodate the daily peak under risk level $\eta$ satisfies
\begin{equation}
\begin{aligned}
    \Pr(y > \text{DASC}_n, Y | \boldsymbol{x}_n) &\leq \eta, \\
    \Rightarrow 
    \Pr(y > \text{DASC}_n | Y, \boldsymbol{x}) 
    \Pr(Y | \boldsymbol{x}_n) &\leq \eta,
\end{aligned}
\end{equation}
where the conditional daily peak density $\Pr(y | Y, \boldsymbol{x}_n)$ is estimated using the proposed approach. As discussed in \autoref{sec:problem_statement}, under worst-case baseline probability $\Pr(Y | \boldsymbol{x}_n) \equiv 1$,
\begin{equation}
\begin{aligned}
    \Pr(y > \text{DASC}_n | Y, \boldsymbol{x})
    &\leq \eta, \\
    \Rightarrow \text{DASC}_n \geq F^{-1} (\alpha;\hat{\boldsymbol{\theta}}_n) &= \text{VaR}_n(\alpha),
\end{aligned}\label{eq:dasc}
\end{equation}
where confidence $\alpha=1 - \eta$ and $F^{-1}(\alpha;\boldsymbol{\theta}_n)$, also known as the \gls{VaR}, is calculated via \autoref{eq:inverse_cdf}. The relationship between the \gls{VaR}, the inverse \gls{CDF}, the \gls{DASC}, and the \gls{LOLP} is displayed in \autoref{fig:quantile_function}. 

\paragraph{Competitor models}

\begin{figure}[htb!]
    \centering
    \begin{minipage}[t]{\textwidth}
        \centering
        \includegraphics[width=\textwidth, page=1]{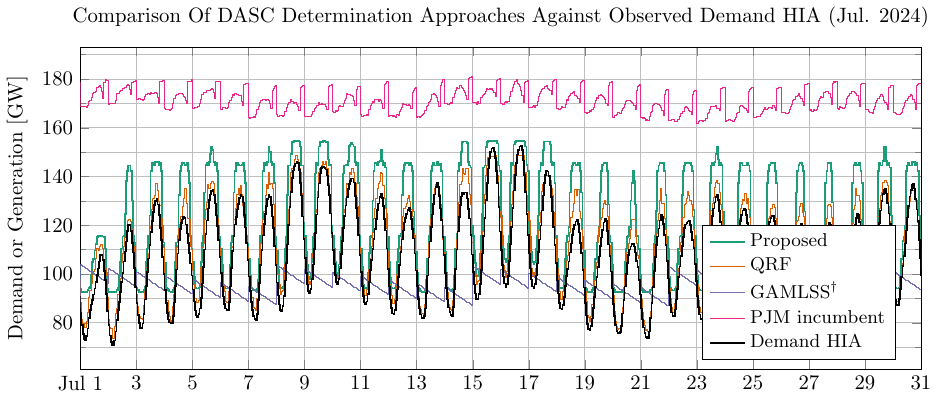}
    \end{minipage}
    \hfill
    \begin{minipage}[t]{\textwidth}
        \centering
        \includegraphics[width=\textwidth, page=2]{999-fig-016.pdf}
    \end{minipage}
    \caption{Comparison of the scheduling capacity determined using the incumbent scheduling practice by PJM, using the proposed approach, using QRF, and using a reduced GAMLSS model in high-risk months of study year 2024. DASC---day-ahead scheduling capacity, HIA---hourly integrated average, QRF---quantile regression forest, GAMLSS---generalized additive model for location, scale, and shape.}
    \label{fig:pjm_testing_prediction}
\end{figure}

The \gls{DASC} determined from the daily risk requirement $\eta$ using the proposed method and \autoref{eq:dasc} changes each hour as conditions change and are compared against the following competitors in \autoref{fig:pjm_testing_prediction}. 

\begin{itemize}
    \item The non-parametric competitor \gls{QRF}, representative of the entire class of quantile-based methods, is fitted on covariate-target pairs (shown in \autoref{fig:pjm_training}) identical to the proposed method to directly predict target quantile $\alpha=1-\eta$ of the daily peak demand per \autoref{eq:nerc_risk}.
    
    \item The parametric competitor \gls{GAMLSS} is fitted on covariate-target pairs using a reduced covariate structure that only includes calendar information. The predicted \gls{GEV} parameters are applied to \autoref{eq:dasc} to produce the \gls{DASC} in the same manner as the proposed method. Numerous attempts at fitting \gls{GAMLSS} on the full covariate in \autoref{fig:pjm_training} failed due to non-convergence. The discussion is expanded in \autoref{sec:gamlss_convergence}. 

    \item The \gls{DASC} resulting from the incumbent PJM scheduling practice is the economic maximum of generating offers in the energy market obtained from PJM Data Miner 2.
\end{itemize}

\begin{figure}[htb!]
    \centering
    \begin{minipage}[t]{\textwidth}
        \centering
        \includegraphics[width=\textwidth, page=1]{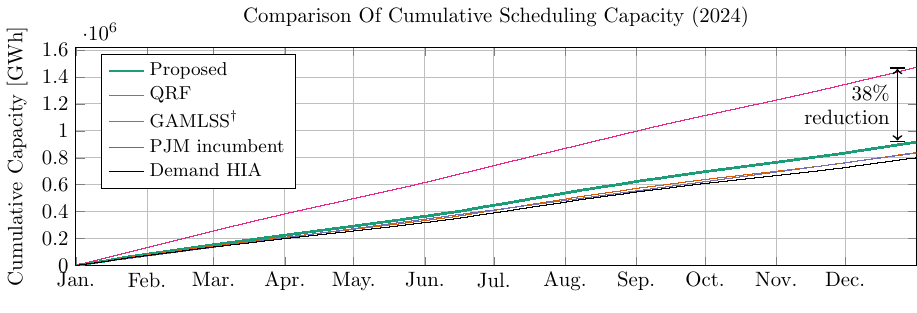}
    \end{minipage}
    \hfill
    \begin{minipage}[t]{\textwidth}
        \centering
        \includegraphics[width=\textwidth, page=2]{999-fig-017.pdf}
    \end{minipage}
    \caption{Comparison of cumulative capacity and retrospective capacity shortfall in 2024. The bottom panel demonstrates systematic failure of QRF and reduced GAMLSS in ensuring resource adequacy. DASC---day-ahead scheduling capacity. QRF---quantile regression forest.}
    \label{fig:pjm_cumulative_capacity}
\end{figure}

It can be observed from \autoref{fig:pjm_testing_prediction} that over the two high-risk months of the year, the \gls{DASC} determined by the proposed method envelopes the demand curve, the PJM incumbent practice appears to over-commit the capacity, and the reduced \gls{GAMLSS} model apparently fails to adequately model peak demand.
The discussion on the reduced \gls{GAMLSS} is expanded in \autoref{sec:gamlss_convergence}. 
While the difference between \gls{QRF} and the proposed method is not obvious in \autoref{fig:pjm_testing_prediction}, systematic failure of \gls{QRF} to ensure resource adequacy is to be demonstrated in the subsequent cumulative capacity analysis.

\paragraph{January 17 capacity increase}
As shown in \autoref{fig:pjm_testing_prediction}, between January 15 and 23, the proposed method predicted an elevated \gls{DASC} despite relatively moderate demand. 
Retrospectively, this increase aligns with historically low temperatures during Winter Storm Gerri, when PJM implemented emergency operating procedures in parts of the system from January 14 to 22, 2024, due to cold weather.
This observation highlights the value of the proposed method as a risk management tool, as it correctly signaled an increase in required capacity in response to the elevated probability of extreme peak demand under low temperature conditions.

\paragraph{Cumulative capacity analysis}
The cumulative capacity over the test year 2024 for the proposed method, \gls{QRF}, \gls{GAMLSS}, and the PJM incumbent scheduling practice is compared against the cumulative demand in \autoref{fig:pjm_cumulative_capacity}. 
The end-of-year cumulative statistics are shown in \autoref{tab:cumulative-statistics}.
The shown cumulative capacity shortfall is the sum of the capacity shortfall over hourly intervals in 2024 in which the demand exceeded the scheduled capacity.
It can be observed that, paradoxically, although \gls{QRF} and the reduced \gls{GAMLSS} produce more optimistic total capacity reduction, they systematically fail to ensure resource adequacy. 
In particular, the cumulative shortfall of the reduced \gls{GAMLSS} increased beyond the vertical scale of \autoref{fig:pjm_cumulative_capacity}.
The proposed approach achieves a 38 percent reduction in total committed capacity while maintaining reliability, as evidenced by zero capacity shortfalls.

\begin{table}[htb!]
    \centering
    \scalebox{0.8}{
    \begin{tabular}{lrrr}
        \hline
        Approach &
        Total Capacity &
        Reduction (\%) &
        Total Shortfall \\
        \hline
        PJM incumbent & \SI{1470573}{\unit{GWh}} & - & 0 \\
        \textbf{Proposed} & \SI{910052}{\unit{GWh}} & $38\%$ & 0 \\
        QRF & \SI{838464}{\unit{GWh}} & $43\%$ & \SI{229}{\unit{GWh}} \\
        GAMLSS\textsuperscript{\textdagger} & \SI{835071}{\unit{GWh}} & $43\%$ & \SI{44048}{\unit{GWh}} \\
        \hline
    \end{tabular}
    }
    \caption{Comparison of cumulative scheduling capacity at the end of test year 2024. The reduction percentage is calculated against the capacity determined using the incumbent scheduling practice by PJM. A nonzero shortfall indicates retrospective capacity inadequacy.}
    \label{tab:cumulative-statistics}
\end{table}

\subsection{Investigation of non-convergence of competitor GAMLSS}
\label{sec:gamlss_convergence}

Although \gls{GAMLSS} was implemented for comparison, the estimation procedure on the full covariate structure in \autoref{fig:pjm_training} failed to converge under the case study conditions. 
Numerous attempts were made to reduce the covariate structure.
In one instance where convergence is achieved, the covariate only includes calendar variables. 
This causes the \gls{GAMLSS} model to neglect the temperature-dependent effects and demand patterns shaped by preceding days on the peak demand density. 
The fitted model was retained for benchmarking in the case study.
However, as demonstrated in the results in \autoref{fig:pjm_testing_prediction} and \autoref{fig:pjm_cumulative_capacity}, even this successfully converged case systematically failed to capture the desired nonstationary behavior of the peak demand. For transparency, the fitting attempts and failure logs are documented in \ref{sec:gamlss_log}.

It is hypothesized that the divergence of \gls{GAMLSS} in the case study arises from numerical issues inherent to the \gls{MLE} method when applied to the \gls{GEV}. Specifically, singularities in the Fisher information $\boldsymbol{I}(\hat{\boldsymbol{\theta}})$ \autoref{eq:fisher} can render the optimization ill-conditioned. Poles in the gamma function $\Gamma(x)$ and the digamma function $\psi(x)$ can destabilize the \gls{NGD} process, preventing convergence. 
While this explanation is consistent with well-documented difficulties of \gls{MLE} for \gls{GEV} in irregular data settings and with the findings of \cite{mayr_generalized_2012}, it should be emphasized that this remains a hypothesis and would need to be formally validated by specialized statistical studies.

\subsection{Unique contribution: expected unserved energy}
The \gls{EUE}, defined as the expected unserved energy due to capacity shortages, is typically reported annually in megawatt hours. The proposed approach enables estimation of \gls{EUE} at a daily resolution. In maximum capacity emergencies, the day-ahead \gls{EUE} can inform the \gls{RTO} of the likely magnitude and duration of load shedding, whether conservation and \gls{DSM} measures suffice, and whether critical loads may be impacted.  Assuming worst-case baseline probability $\Pr(Y | \boldsymbol{x}_n)\equiv1$ where condition $\boldsymbol{x}_n$ of hour $n$ always results in the daily peak, the daily \gls{EUE} under the \gls{DASC} determined from \autoref{eq:dasc} based on confidence $\alpha=1 - \eta$ is given by
\begin{equation}
\begin{aligned}
    \text{EUE} 
    &= (1 - \alpha) \sum_{n=1}^{24}  \left(
        \text{CVaR}_n(\alpha)
        - \text{VaR}_n(\alpha) 
    \right),
\end{aligned}\label{eq:eue}
\end{equation}
and for lower incomplete gamma function $\Gamma_l(s,x)=\int_0^x t^{s-1} \exp(-t) dt$, logarithmic integral function $\text{li}(x)=\int_0^x dx/\log x$, the \gls{CVaR} is given by
\begin{equation}
\text{CVaR}(\alpha) =
\begin{cases}
    \mu + \frac{\sigma}{1 - \alpha} 
    \left[
        \gamma_\text{EM} - 
        \text{li}(\alpha) + 
        \alpha\log(-\log\alpha) 
    \right] & \xi = 0,\\
    \mu + \frac{\sigma}{(1-\alpha)\xi} \left[
        \Gamma_l(1 - \xi, -\log\alpha) - (1-\alpha)
    \right] & \xi \neq 0.
\end{cases}
\end{equation}

\begin{figure}[htb!]
    \centering
    \includegraphics[width=\textwidth]{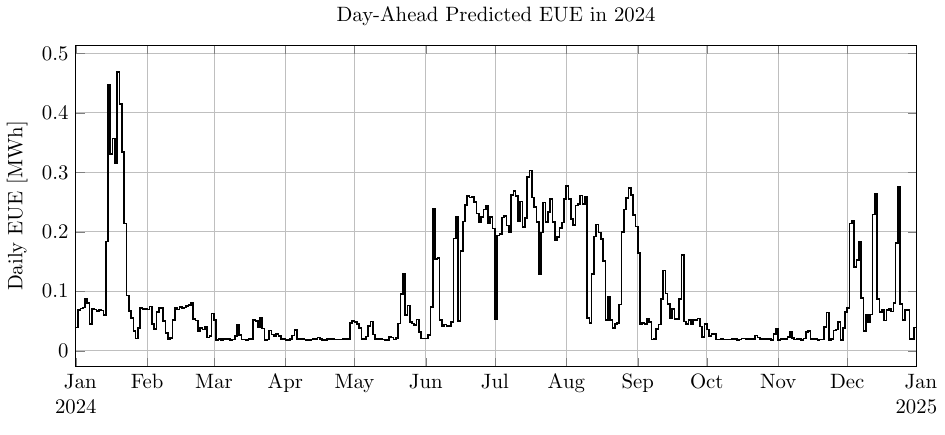}
    \caption{Predicted daily EUE in study year 2024. EUE---expected unserved energy.}
    \label{fig:pjm_eue}
\end{figure}

The \gls{EUE} for test year 2024, calculated from \autoref{eq:eue} and shown in \autoref{fig:pjm_eue}, indicates that severe \gls{EUE} magnitudes occur in the winter, while summer contributes to cumulative totals. 
Under the \gls{LOLP}-based \gls{DASC} in \autoref{eq:nerc_risk}, summing daily values provides an annual \gls{EUE} of $\SI{32.37}{\mega\watt\hour}$, or $\SI{0.04}{\unit{PPM}}$ of the $\SI{799269.28}{\giga\watt\hour}$ total demand.
In contrast, the 2024 \gls{NERC} probabilistic assessment report determines that the \gls{EUE} of PJM in upcoming years is $\SI{0.00}{\unit{PPM}}$ \cite{nerc_north_american_electric_reliability_corporation_2024_2024}, and the report shows \gls{EUE} in the range of $\SI{0}{\unit{PPM}} - \SI{100}{\unit{PPM}}$ for a typical \gls{RTO}.

    \section{Conclusion}
This work develops an ensemble-based machine learning framework that integrates tree-based learners with extreme value analysis to forecast nonstationary peak demand distributions, addressing a critical gap in short-term resource adequacy planning under growing system volatility. Validation on the PJM interconnection shows that the method can reduce committed capacity by 38 percent, translating into substantial economic and operational benefits. The originality of the contribution lies in uniting extreme value analysis with machine learning to overcome the convergence and functional-form limitations of existing nonstationary density models. Despite these advances, challenges remain in hyperparameter tuning, goodness-of-fit validation, and the need for large datasets to capture rare events. Future research should integrate advanced deep learning for improved temporal accuracy, further evaluate the framework for maturity, and explore evolving infrastructures with high renewable penetration and across other critical engineering disciplines. By situating this methodology within an international context, the work establishes a foundation for more resilient and economically efficient energy systems worldwide.

    \section*{Acknowledgements}
\begin{itemize}
\item This paper is partly derived from the work supported by the U.S. Department of Energy’s Office of Energy Efficiency and Renewable Energy (EERE) through the Solar Energy Technology Office (SETO) under Award DE-EE0009357. We thank the agency for its financial support. The funding source has no involvement in
the development of the methodology, design, preparation, or interpretation of the case study, or the decision to submit for publication.
\item We would like to thank Haley Northrup for her support in improving this paper's organization, wording, clarity, and impact-related aspects.
\end{itemize}

    \section*{Declaration of generative AI and AI-assisted technologies in the writing process}
During the preparation of this work, the authors used generative \gls{AI} tools, including ChatGPT, ScholarGPT, and Consensus, to assist in identifying relevant academic and industry publications, and NotebookLM for indexing and organizing publicly available training materials, presentations, manuals, and white papers from PJM, ISO New England, and \gls{NERC}. 
During the revision, ChatGPT was employed to improve the clarity, conciseness, and readability in terms of grammar, style, and formatting. 
While some statements were informed by \gls{AI}-generated text, information provided by these tools was critically examined, substantiated, and rigorously vetted by the authors.
The authors take full responsibility for the content of the published article.

    \appendix
\section{GEV gradient components and Fisher information}
\label{sec:gev_gradient_formulas}

For completeness, the formulas for the ordinary gradient elements and Fisher information matrix used in \gls{GEV} maximum likelihood estimation are reproduced. The ordinary gradient $\nabla L$ is derived by \cite{dey_extreme_2016}:
\begin{subequations}\label{eq:ordinary_gradient}
\begin{align}
\frac{\partial L}{\partial\mu} &= 
\frac{1}{\sigma\omega}(1 + \xi - \nu), \\
\frac{\partial L}{\partial\sigma} &=
\frac{1}{\sigma} \left[-1 + \frac{1}{\sigma \omega} (1 + \xi - \nu)(y - \mu) \right], \\
\frac{\partial L}{\partial\xi} &= 
\frac{1}{\xi^2} \log \left( \omega (1 - \nu) \right) - \frac{1}{\sigma\omega} (1 + \xi - \nu)(y - \mu)\xi, \\
\omega &= 1 + \frac{\xi}{\sigma}(y - \mu),\\
\nu &= \omega^{-\frac{1}{\xi}}.
\end{align}
\end{subequations}

For gamma function $\Gamma(x) = \int_0^\infty \tau^{x-1} \exp (-\tau) d\tau$, digamma function $\psi(x) = \partial\log\Gamma(x)/ \partial x$, and Euler--Mascheroni constant $\gamma_\text{EM} = -\psi(1) \approx 0.5772$, Fisher information $\boldsymbol{I}(\boldsymbol{\theta})$ is derived by \cite{prescott_maximum_1980}:
\begin{subequations}\label{eq:fisher}
\begin{align}
I_{\mu\mu} &= \frac{1}{\sigma^2} p, \\
I_{\mu\sigma} &= -\frac{1}{\sigma^2\xi} \left( p - \Gamma(2+\xi) \right), \\
I_{\mu\xi} &= \frac{1}{\sigma\xi} \left( q - \frac{p}{\xi} \right), \\
I_{\sigma\sigma} &= \frac{1}{\sigma^2\xi^2} \left( 1 - 2\Gamma(2+\xi) + p \right), \\
I_{\sigma\xi} &= \frac{1}{\sigma\xi^2} \biggl(\frac{1}{\xi} (1 - \Gamma(2+\xi) + p) + 1 - \gamma_\text{EM} - q \biggr), \\
I_{\xi\xi} &= \frac{1}{\xi^2} \left(\frac{\pi^2}{6} + 
    \left(1 - \gamma_\text{EM} + \frac{1}{\xi}\right)^2 - \frac{2q}{\xi} + \frac{p}{\xi^2}\right), \\
p &= (1 + \xi)^2 \Gamma(1 + 2\xi), \\
q &= \Gamma(2 + \xi) \left(\psi(1 + \xi) + \frac{1 + \xi}{\xi}\right).
\end{align}
\end{subequations}

\section{GAMLSS program listings in the PJM study}
\label{sec:gamlss_log}

In \autoref{lst:pjm_gamlss_full_covariate}, the covariate for \gls{GAMLSS} is the same as that used by the proposed method in the case study \autoref{sec:case_study}. 
The effect of temperature and the PJM base forecast on the peak demand density is modeled by penalized B-splines. Day, day of week, month, and hour are treated as categorical variables. As described in the main text, the responsibility of modeling autocorrelation is delegated to the PJM base forecast.
The output of the program is shown in \autoref{lst:pjm_gamlss_full_covariate_output}, showing no convergence.
In a subsequent attempt, the covariate element \texttt{Forecast.MW} is removed, and the output of the program is similar to \autoref{lst:pjm_gamlss_full_covariate_output}, showing no convergence.
In a third attempt, in addition to removing \texttt{Forecast.MW}, the covariate element \texttt{Degrees.F} is also removed from the model, and the program is shown in \autoref{lst:pjm_gamlss_covariate_working}.
The program converges. As discussed in the main text in \autoref{fig:pjm_testing_prediction}, the deletion of covariate elements from the \gls{GAMLSS} model causes the predictions to be based on only calendar information, which neglects the temperature-dependent effects and demand patterns shaped by preceding days on the peak demand density. 

\begin{lstlisting}[language=r, 
caption={GAMLSS program in PJM study, full covariate},
label={lst:pjm_gamlss_full_covariate}]
#!/usr/bin/env Rscript
library(gamlss)
library(gamlssx)

train <- read.csv('datasets/pjm/peak_training.csv')
test <- read.csv('datasets/pjm/whole_testing.csv')

# Drop the first colummn (Timestamp)
train_no_timestamps <- subset(train, select = -1)
test_no_timestamps <- subset(test, select = -1)

model <- fitGEV(
    Load.MW ~ pb(Forecast.MW) + pb(Degrees.F) + Day + DoW + Month + Hour,
    sigma.fo = ~ pb(Forecast.MW) + pb(Degrees.F) + Day + DoW + Month + Hour,
    nu.fo = ~ pb(Forecast.MW) + pb(Degrees.F) + Day + DoW + Month + Hour,
    data=train_no_timestamps
)

mu_hat    <- predict(model, what = "mu",  newdata = test_no_timestamps)
sigma_hat <- predict(model, what = "sigma", newdata = test_no_timestamps)
xi_hat    <- predict(model, what = "nu", newdata = test_no_timestamps)

parameter_estimates <- data.frame(index=test$Time,
    mu_hat=mu_hat,
    sigma_hat=sigma_hat,
    xi_hat=xi_hat)

write.csv(
    parameter_estimates,
    file='192-run_competitor2_on_pjm.csv',
    quote=FALSE,
    row.names=FALSE
)
\end{lstlisting}

\begin{lstlisting}[style=plainoutput, caption={Output of GAMLSS program in PJM study, full covariate},
label={lst:pjm_gamlss_full_covariate_output}]
Loading required package: splines
Loading required package: gamlss.data

Attaching package: 'gamlss.data'

The following object is masked from 'package:datasets':

    sleep

Loading required package: gamlss.dist
Loading required package: nlme
Loading required package: parallel
 **********   GAMLSS Version 5.4-22  ********** 
For more on GAMLSS look at https://www.gamlss.com/
Type gamlssNews() to see new features/changes/bug fixes.

GAMLSS-RS iteration 1: Global Deviance = 27913.59 
GAMLSS-RS iteration 1: Global Deviance = 27911.48 
GAMLSS-RS iteration 1: Global Deviance = 27914.85 
Error in fitGEV(Load.MW ~ pb(Forecast.MW) + pb(Degrees.F) + Day + DoW +  : 
  No convergence. An error was thrown from the last call to gamlss()
In addition: There were 50 or more warnings (use warnings() to see the first 50)
Execution halted
\end{lstlisting}

\begin{lstlisting}[language=R, 
caption={GAMLSS program in PJM study, calendar-only covariate. Lines identical to \autoref{lst:pjm_gamlss_full_covariate} are elided.},
label={lst:pjm_gamlss_covariate_working}]
...

model <- fitGEV(
    Load.MW ~ Day + DoW + Month + Hour,
    sigma.fo = ~ Day + DoW + Month + Hour,
    nu.fo = ~ Day + DoW + Month + Hour,
    data=train_no_timestamps
)

...
\end{lstlisting}

    \bibliographystyle{elsarticle-num}
    \bibliography{900-references}

\begin{thebibliography}{10}
\expandafter\ifx\csname url\endcsname\relax
  \def\url#1{\texttt{#1}}\fi
\expandafter\ifx\csname urlprefix\endcsname\relax\def\urlprefix{URL }\fi
\expandafter\ifx\csname href\endcsname\relax
  \def\href#1#2{#2} \def\path#1{#1}\fi

\bibitem{flores_2021_2023}
N.~M. Flores, H.~McBrien, V.~Do, M.~V. Kiang, J.~Schlegelmilch, J.~A. Casey, \href{https://www.nature.com/articles/s41370-022-00462-5}{The 2021 texas power crisis: distribution, duration, and disparities}, Journal of Exposure Science \& Environmental Epidemiology 33~(1) (2023) 21--31.
\newblock \href {https://doi.org/10.1038/s41370-022-00462-5} {\path{doi:10.1038/s41370-022-00462-5}}.
\newline\urlprefix\url{https://www.nature.com/articles/s41370-022-00462-5}

\bibitem{north_american_electric_reliability_corporation_bal-502-rf-03_2025}
{North American Electric Reliability Corporation}, \href{https://www.nerc.com/pa/Stand/Reliability%20Standards/BAL-502-RF-03.pdf}{{BAL}-502-{RF}-03: planning resource adequacy analysis, assessment and documentation}, Tech. rep., North American Electric Reliability Corporation, United States (Sep. 2025).
\newline\urlprefix\url{https://www.nerc.com/pa/Stand/Reliability%20Standards/BAL-502-RF-03.pdf}

\bibitem{australian_energy_market_commission_national_2025}
{Australian Energy Market Commission}, \href{https://energy-rules.aemc.gov.au/ner/674/657539#3.9.3C}{National electricity rules -- clause 3.9.3c: reliability standard and interim reliability measure}, Tech. rep., Australian Energy Market Commission, Australia (Sep. 2025).
\newline\urlprefix\url{https://energy-rules.aemc.gov.au/ner/674/657539#3.9.3C}

\bibitem{european_parliament_council_of_the_european_union_regulation_2025}
{European Parliament; Council of the European Union}, \href{https://eur-lex.europa.eu/eli/reg/2019/941/oj}{Regulation ({EU}) 2019/941 of the european parliament and of the council of 5 {June} 2019 on risk-preparedness in the electricity sector and repealing directive 2005/89/{EC}}, Official Journal of the European Union L 158~(1) (2025) 1--21, place: European Union.
\newline\urlprefix\url{https://eur-lex.europa.eu/eli/reg/2019/941/oj}

\bibitem{iso_new_england_operational_2025}
{ISO New England}, \href{https://www.iso-ne.com/static-assets/documents/2023/05/a10_operational_impact_of_extreme_weather_events.pdf}{Operational impact of extreme weather events: preliminary results of energy adequacy studies for winter 2027}, Tech. rep., ISO New England, United States (Sep. 2025).
\newline\urlprefix\url{https://www.iso-ne.com/static-assets/documents/2023/05/a10_operational_impact_of_extreme_weather_events.pdf}

\bibitem{bousquet_extreme_2021}
N.~Bousquet, P.~Bernardara (Eds.), \href{https://link.springer.com/10.1007/978-3-030-74942-2}{Extreme value theory with applications to natural gazards: from statistical theory to industrial practice}, Springer International Publishing, Cham, 2021.
\newblock \href {https://doi.org/10.1007/978-3-030-74942-2} {\path{doi:10.1007/978-3-030-74942-2}}.
\newline\urlprefix\url{https://link.springer.com/10.1007/978-3-030-74942-2}

\bibitem{hassan_integrated_2024}
M.~Hassan, A.~M. El-Rifaie, M.~Beshr, E.~Beshr, \href{https://ieeexplore.ieee.org/document/10616114/}{Integrated smart risk management for {Siwa} solar energy systems: a case study and strategies}, IEEE Access 12 (2024) 106025--106041.
\newblock \href {https://doi.org/10.1109/ACCESS.2024.3436018} {\path{doi:10.1109/ACCESS.2024.3436018}}.
\newline\urlprefix\url{https://ieeexplore.ieee.org/document/10616114/}

\bibitem{bollerslev_generalized_1986}
T.~Bollerslev, \href{https://linkinghub.elsevier.com/retrieve/pii/0304407686900631}{Generalized autoregressive conditional heteroskedasticity}, Journal of Econometrics 31~(3) (1986) 307--327.
\newblock \href {https://doi.org/10.1016/0304-4076(86)90063-1} {\path{doi:10.1016/0304-4076(86)90063-1}}.
\newline\urlprefix\url{https://linkinghub.elsevier.com/retrieve/pii/0304407686900631}

\bibitem{rigby_generalized_2005}
R.~A. Rigby, D.~M. Stasinopoulos, \href{https://academic.oup.com/jrsssc/article/54/3/507/7113027}{Generalized additive models for location, scale and shape}, Journal of the Royal Statistical Society Series C: Applied Statistics 54~(3) (2005) 507--554.
\newblock \href {https://doi.org/10.1111/j.1467-9876.2005.00510.x} {\path{doi:10.1111/j.1467-9876.2005.00510.x}}.
\newline\urlprefix\url{https://academic.oup.com/jrsssc/article/54/3/507/7113027}

\bibitem{debele_around_2017}
S.~E. Debele, E.~Bogdanowicz, W.~G. Strupczewski, \href{http://link.springer.com/10.1007/s11600-017-0072-3}{Around and about an application of the {GAMLSS} package to non-stationary flood frequency analysis}, Acta Geophysica 65~(4) (2017) 885--892.
\newblock \href {https://doi.org/10.1007/s11600-017-0072-3} {\path{doi:10.1007/s11600-017-0072-3}}.
\newline\urlprefix\url{http://link.springer.com/10.1007/s11600-017-0072-3}

\bibitem{koenker_regression_1978}
R.~Koenker, G.~Bassett, \href{https://www.jstor.org/stable/1913643?origin=crossref}{Regression quantiles}, Econometrica 46~(1) (1978) 33.
\newblock \href {https://doi.org/10.2307/1913643} {\path{doi:10.2307/1913643}}.
\newline\urlprefix\url{https://www.jstor.org/stable/1913643?origin=crossref}

\bibitem{guo_probabilistic_2024}
H.~Guo, B.~Huang, J.~Wang, \href{https://linkinghub.elsevier.com/retrieve/pii/S2666792424000039}{Probabilistic load forecasting for integrated energy systems using attentive quantile regression temporal convolutional network}, Advances in Applied Energy 14 (2024) 100165.
\newblock \href {https://doi.org/10.1016/j.adapen.2024.100165} {\path{doi:10.1016/j.adapen.2024.100165}}.
\newline\urlprefix\url{https://linkinghub.elsevier.com/retrieve/pii/S2666792424000039}

\bibitem{li_probabilistic_2024}
G.~Li, R.~Zhang, S.~Bu, J.~Zhang, J.~Gao, \href{https://linkinghub.elsevier.com/retrieve/pii/S0142061524004216}{Probabilistic prediction-based multi-objective optimization approach for multi-energy virtual power plant}, International Journal of Electrical Power \& Energy Systems 161 (2024) 110200.
\newblock \href {https://doi.org/10.1016/j.ijepes.2024.110200} {\path{doi:10.1016/j.ijepes.2024.110200}}.
\newline\urlprefix\url{https://linkinghub.elsevier.com/retrieve/pii/S0142061524004216}

\bibitem{zuege_wind_2025}
C.~V. Zuege, S.~F. Stefenon, C.~K. Yamaguchi, V.~C. Mariani, G.~V. Gonzalez, L.~D.~S. Coelho, \href{https://linkinghub.elsevier.com/retrieve/pii/S0142061525002510}{Wind speed forecasting approach using conformal prediction and feature importance selection}, International Journal of Electrical Power \& Energy Systems 168 (2025) 110700.
\newblock \href {https://doi.org/10.1016/j.ijepes.2025.110700} {\path{doi:10.1016/j.ijepes.2025.110700}}.
\newline\urlprefix\url{https://linkinghub.elsevier.com/retrieve/pii/S0142061525002510}

\bibitem{parejo_probabilistic_2025}
A.~Parejo, S.~García, E.~Personal, J.~I. Guerrero, A.~Carrasco, a.~et, \href{https://ieeexplore.ieee.org/document/10433161/}{Probabilistic forecasting framework oriented to distribution networks and microgrids}, IEEE Transactions on Automation Science and Engineering 22 (2025) 1183--1195.
\newblock \href {https://doi.org/10.1109/TASE.2024.3361651} {\path{doi:10.1109/TASE.2024.3361651}}.
\newline\urlprefix\url{https://ieeexplore.ieee.org/document/10433161/}

\bibitem{botman_global_2025}
L.~Botman, J.~Lago, T.~Becker, K.~Vanthournout, B.~D. Moor, \href{https://linkinghub.elsevier.com/retrieve/pii/S0306261924025522}{A global probabilistic approach for short-term forecasting of individual households electricity consumption}, Applied Energy 382 (2025) 125168.
\newblock \href {https://doi.org/10.1016/j.apenergy.2024.125168} {\path{doi:10.1016/j.apenergy.2024.125168}}.
\newline\urlprefix\url{https://linkinghub.elsevier.com/retrieve/pii/S0306261924025522}

\bibitem{hu_probabilistic_2024}
J.~Hu, W.~Hu, D.~Cao, X.~Sun, J.~Chen, Y.~Huang, et~al., \href{https://linkinghub.elsevier.com/retrieve/pii/S0960148124003185}{Probabilistic net load forecasting based on transformer network and {Gaussian} process-enabled residual modeling learning method}, Renewable Energy 225 (2024) 120253.
\newblock \href {https://doi.org/10.1016/j.renene.2024.120253} {\path{doi:10.1016/j.renene.2024.120253}}.
\newline\urlprefix\url{https://linkinghub.elsevier.com/retrieve/pii/S0960148124003185}

\bibitem{zhu_peak_2024}
N.~Zhu, Y.~Wang, K.~Yuan, J.~Lv, B.~Su, K.~Zhang, \href{https://linkinghub.elsevier.com/retrieve/pii/S0142061524005635}{Peak interval-focused wind power forecast with dynamic ramp considerations}, International Journal of Electrical Power \& Energy Systems 163 (2024) 110340.
\newblock \href {https://doi.org/10.1016/j.ijepes.2024.110340} {\path{doi:10.1016/j.ijepes.2024.110340}}.
\newline\urlprefix\url{https://linkinghub.elsevier.com/retrieve/pii/S0142061524005635}

\bibitem{marulanda_modeling_2024}
G.~Marulanda, A.~Bello, J.~Reneses, \href{https://linkinghub.elsevier.com/retrieve/pii/S0142061524001108}{Modeling wind energy imbalance risk in medium-term generation planning models: a methodological proposal for large scale applications}, International Journal of Electrical Power \& Energy Systems 157 (2024) 109889.
\newblock \href {https://doi.org/10.1016/j.ijepes.2024.109889} {\path{doi:10.1016/j.ijepes.2024.109889}}.
\newline\urlprefix\url{https://linkinghub.elsevier.com/retrieve/pii/S0142061524001108}

\bibitem{xu_interpretable_2024}
C.~Xu, G.~Chen, \href{https://linkinghub.elsevier.com/retrieve/pii/S0142061523005720}{Interpretable transformer-based model for probabilistic short-term forecasting of residential net load}, International Journal of Electrical Power \& Energy Systems 155 (2024) 109515.
\newblock \href {https://doi.org/10.1016/j.ijepes.2023.109515} {\path{doi:10.1016/j.ijepes.2023.109515}}.
\newline\urlprefix\url{https://linkinghub.elsevier.com/retrieve/pii/S0142061523005720}

\bibitem{zhu_ultra-short-term_2024}
J.~Zhu, Y.~He, X.~Yang, S.~Yang, \href{https://linkinghub.elsevier.com/retrieve/pii/S0196890424000037}{Ultra-short-term wind power probabilistic forecasting based on an evolutionary non-crossing multi-output quantile regression deep neural network}, Energy Conversion and Management 301 (2024) 118062.
\newblock \href {https://doi.org/10.1016/j.enconman.2024.118062} {\path{doi:10.1016/j.enconman.2024.118062}}.
\newline\urlprefix\url{https://linkinghub.elsevier.com/retrieve/pii/S0196890424000037}

\bibitem{selcen_ayaz_probabilistic_2024}
M.~Selcen~Ayaz, M.~Malekpour, R.~Azizipanah-Abarghooee, M.~Karimi, V.~Terzija, \href{https://linkinghub.elsevier.com/retrieve/pii/S0142061524002370}{Probabilistic photovoltaic generation and load demand uncertainties modelling for active distribution networks hosting capacity calculations}, International Journal of Electrical Power \& Energy Systems 159 (2024) 110016.
\newblock \href {https://doi.org/10.1016/j.ijepes.2024.110016} {\path{doi:10.1016/j.ijepes.2024.110016}}.
\newline\urlprefix\url{https://linkinghub.elsevier.com/retrieve/pii/S0142061524002370}

\bibitem{li_probabilistic_2023}
D.~Li, Y.~Tan, Y.~Zhang, S.~Miao, S.~He, \href{https://linkinghub.elsevier.com/retrieve/pii/S0142061522007396}{Probabilistic forecasting method for mid-term hourly load time series based on an improved temporal fusion transformer model}, International Journal of Electrical Power \& Energy Systems 146 (2023) 108743.
\newblock \href {https://doi.org/10.1016/j.ijepes.2022.108743} {\path{doi:10.1016/j.ijepes.2022.108743}}.
\newline\urlprefix\url{https://linkinghub.elsevier.com/retrieve/pii/S0142061522007396}

\bibitem{he_short-term_2022}
Y.~He, J.~Xiao, X.~An, C.~Cao, J.~Xiao, \href{https://linkinghub.elsevier.com/retrieve/pii/S0142061522002721}{Short-term power load probability density forecasting based on {GLRQ}-stacking ensemble learning method}, International Journal of Electrical Power \& Energy Systems 142 (2022) 108243.
\newblock \href {https://doi.org/10.1016/j.ijepes.2022.108243} {\path{doi:10.1016/j.ijepes.2022.108243}}.
\newline\urlprefix\url{https://linkinghub.elsevier.com/retrieve/pii/S0142061522002721}

\bibitem{gruosso_probabilistic_2019}
G.~Gruosso, P.~Maffezzoni, Z.~Zhang, L.~Daniel, \href{https://linkinghub.elsevier.com/retrieve/pii/S0142061518307671}{Probabilistic load flow methodology for distribution networks including loads uncertainty}, International Journal of Electrical Power \& Energy Systems 106 (2019) 392--400.
\newblock \href {https://doi.org/10.1016/j.ijepes.2018.10.023} {\path{doi:10.1016/j.ijepes.2018.10.023}}.
\newline\urlprefix\url{https://linkinghub.elsevier.com/retrieve/pii/S0142061518307671}

\bibitem{waldmann_quantile_2018}
E.~Waldmann, \href{https://journals.sagepub.com/doi/10.1177/1471082X18759142}{Quantile regression: a short story on how and why}, Statistical Modelling 18~(3-4) (2018) 203--218.
\newblock \href {https://doi.org/10.1177/1471082X18759142} {\path{doi:10.1177/1471082X18759142}}.
\newline\urlprefix\url{https://journals.sagepub.com/doi/10.1177/1471082X18759142}

\bibitem{rosenblatt_remarks_1956}
M.~Rosenblatt, \href{http://projecteuclid.org/euclid.aoms/1177728190}{Remarks on some nonparametric estimates of a density function}, The Annals of Mathematical Statistics 27~(3) (1956) 832--837.
\newblock \href {https://doi.org/10.1214/aoms/1177728190} {\path{doi:10.1214/aoms/1177728190}}.
\newline\urlprefix\url{http://projecteuclid.org/euclid.aoms/1177728190}

\bibitem{ertugrul_forecasting_2016}
O.~F. Ertugrul, \href{https://linkinghub.elsevier.com/retrieve/pii/S0142061515005578}{Forecasting electricity load by a novel recurrent extreme learning machines approach}, International Journal of Electrical Power \& Energy Systems 78 (2016) 429--435.
\newblock \href {https://doi.org/10.1016/j.ijepes.2015.12.006} {\path{doi:10.1016/j.ijepes.2015.12.006}}.
\newline\urlprefix\url{https://linkinghub.elsevier.com/retrieve/pii/S0142061515005578}

\bibitem{hochreiter_long_1997}
S.~Hochreiter, J.~Schmidhuber, \href{https://direct.mit.edu/neco/article/9/8/1735-1780/6109}{Long short-term memory}, Neural Computation 9~(8) (1997) 1735--1780.
\newblock \href {https://doi.org/10.1162/neco.1997.9.8.1735} {\path{doi:10.1162/neco.1997.9.8.1735}}.
\newline\urlprefix\url{https://direct.mit.edu/neco/article/9/8/1735-1780/6109}

\bibitem{bai_empirical_2018}
S.~Bai, J.~Z. Kolter, V.~Koltun, \href{https://arxiv.org/abs/1803.01271}{An empirical evaluation of generic convolutional and recurrent networks for sequence modeling}, version Number: 2 (2018).
\newblock \href {https://doi.org/10.48550/ARXIV.1803.01271} {\path{doi:10.48550/ARXIV.1803.01271}}.
\newline\urlprefix\url{https://arxiv.org/abs/1803.01271}

\bibitem{castillo_extreme_2012}
E.~Castillo, \href{https://doi.org/10.1016/C2009-0-22169-6}{Extreme value theory in engineering}, Statistical {Modeling} and {Decision} {Science}, Academic Press, 2012.
\newline\urlprefix\url{https://doi.org/10.1016/C2009-0-22169-6}

\bibitem{majumdar_extreme_2020}
S.~N. Majumdar, A.~Pal, G.~Schehr, \href{https://linkinghub.elsevier.com/retrieve/pii/S0370157319303291}{Extreme value statistics of correlated random variables: a pedagogical review}, Physics Reports 840 (2020) 1--32.
\newblock \href {https://doi.org/10.1016/j.physrep.2019.10.005} {\path{doi:10.1016/j.physrep.2019.10.005}}.
\newline\urlprefix\url{https://linkinghub.elsevier.com/retrieve/pii/S0370157319303291}

\bibitem{leadbetter_extremes_1983}
M.~R. Leadbetter, G.~Lindgren, H.~Rootzén, \href{http://link.springer.com/10.1007/978-1-4612-5449-2}{Extremes and related properties of random sequences and processes}, Springer {Series} in {Statistics}, Springer New York, New York, NY, 1983.
\newblock \href {https://doi.org/10.1007/978-1-4612-5449-2} {\path{doi:10.1007/978-1-4612-5449-2}}.
\newline\urlprefix\url{http://link.springer.com/10.1007/978-1-4612-5449-2}

\bibitem{ganger_statistical_2014}
D.~Ganger, J.~Zhang, V.~Vittal, \href{http://ieeexplore.ieee.org/document/6786040/}{Statistical characterization of wind power ramps via extreme value analysis}, IEEE Transactions on Power Systems 29~(6) (2014) 3118--3119.
\newblock \href {https://doi.org/10.1109/TPWRS.2014.2315491} {\path{doi:10.1109/TPWRS.2014.2315491}}.
\newline\urlprefix\url{http://ieeexplore.ieee.org/document/6786040/}

\bibitem{wang_high_2018}
Z.~Wang, Z.~Liu, Y.~Sun, W.~Gao, C.~Gu, \href{https://ieeexplore.ieee.org/document/8582202/}{High impact low frequency peak load analysis using extreme value theory}, in: 2018 2nd {IEEE} {Conference} on {Energy} {Internet} and {Energy} {System} {Integration} ({EI2}), IEEE, Beijing, 2018, pp. 1--6.
\newblock \href {https://doi.org/10.1109/EI2.2018.8582202} {\path{doi:10.1109/EI2.2018.8582202}}.
\newline\urlprefix\url{https://ieeexplore.ieee.org/document/8582202/}

\bibitem{kullback_information_1951}
S.~Kullback, R.~A. Leibler, \href{http://projecteuclid.org/euclid.aoms/1177729694}{On information and sufficiency}, The Annals of Mathematical Statistics 22~(1) (1951) 79--86.
\newblock \href {https://doi.org/10.1214/aoms/1177729694} {\path{doi:10.1214/aoms/1177729694}}.
\newline\urlprefix\url{http://projecteuclid.org/euclid.aoms/1177729694}

\bibitem{breiman_bagging_1996}
L.~Breiman, \href{http://link.springer.com/10.1007/BF00058655}{Bagging predictors}, Machine Learning 24~(2) (1996) 123--140.
\newblock \href {https://doi.org/10.1007/BF00058655} {\path{doi:10.1007/BF00058655}}.
\newline\urlprefix\url{http://link.springer.com/10.1007/BF00058655}

\bibitem{amari_natural_1998}
S.-i. Amari, \href{https://direct.mit.edu/neco/article/10/2/251-276/6143}{Natural gradient works efficiently in learning}, Neural Computation 10~(2) (1998) 251--276.
\newblock \href {https://doi.org/10.1162/089976698300017746} {\path{doi:10.1162/089976698300017746}}.
\newline\urlprefix\url{https://direct.mit.edu/neco/article/10/2/251-276/6143}

\bibitem{duan_ngboost_2020}
T.~Duan, A.~Avati, D.~Y. Ding, K.~K. Thai, S.~Basu, A.~Ng, et~al., \href{https://proceedings.mlr.press/v119/duan20a.html}{{NGBoost}: natural gradient boosting for probabilistic prediction}, Vol. 119, PMLR, Virtual Event, 2020, pp. 2690--2700.
\newline\urlprefix\url{https://proceedings.mlr.press/v119/duan20a.html}

\bibitem{dey_extreme_2016}
D.~K. Dey, J.~Yan (Eds.), \href{https://www.taylorfrancis.com/books/9781498701310}{Extreme value modeling and risk analysis: methods and applications}, 0th Edition, Chapman and Hall/CRC, 2016.
\newblock \href {https://doi.org/10.1201/b19721} {\path{doi:10.1201/b19721}}.
\newline\urlprefix\url{https://www.taylorfrancis.com/books/9781498701310}

\bibitem{prescott_maximum_1980}
P.~Prescott, A.~T. Walden, \href{https://academic.oup.com/biomet/article-lookup/doi/10.1093/biomet/67.3.723}{Maximum likelihood estimation of the parameters of the generalized extreme-value distribution}, Biometrika 67~(3) (1980) 723--724.
\newblock \href {https://doi.org/10.1093/biomet/67.3.723} {\path{doi:10.1093/biomet/67.3.723}}.
\newline\urlprefix\url{https://academic.oup.com/biomet/article-lookup/doi/10.1093/biomet/67.3.723}

\bibitem{hosking_estimation_1985}
J.~R.~M. Hosking, J.~R. Wallis, E.~F. Wood, \href{http://www.tandfonline.com/doi/abs/10.1080/00401706.1985.10488049}{Estimation of the generalized extreme-value distribution by the method of probability-weighted moments}, Technometrics 27~(3) (1985) 251--261.
\newblock \href {https://doi.org/10.1080/00401706.1985.10488049} {\path{doi:10.1080/00401706.1985.10488049}}.
\newline\urlprefix\url{http://www.tandfonline.com/doi/abs/10.1080/00401706.1985.10488049}

\bibitem{martins_generalized_2000}
E.~S. Martins, J.~R. Stedinger, \href{https://agupubs.onlinelibrary.wiley.com/doi/10.1029/1999WR900330}{Generalized maximum‐likelihood generalized extreme‐value quantile estimators for hydrologic data}, Water Resources Research 36~(3) (2000) 737--744.
\newblock \href {https://doi.org/10.1029/1999WR900330} {\path{doi:10.1029/1999WR900330}}.
\newline\urlprefix\url{https://agupubs.onlinelibrary.wiley.com/doi/10.1029/1999WR900330}

\bibitem{coles_introduction_2001}
S.~Coles, \href{http://link.springer.com/10.1007/978-1-4471-3675-0}{An {Introduction} to statistical modeling of extreme values}, Springer {Series} in {Statistics}, Springer London, London, 2001.
\newblock \href {https://doi.org/10.1007/978-1-4471-3675-0} {\path{doi:10.1007/978-1-4471-3675-0}}.
\newline\urlprefix\url{http://link.springer.com/10.1007/978-1-4471-3675-0}

\bibitem{breiman_classification_2017}
L.~Breiman, J.~H. Friedman, R.~A. Olshen, C.~J. Stone, \href{https://www.taylorfrancis.com/books/9781351460491}{Classification and regression trees}, 1st Edition, Routledge, 2017.
\newblock \href {https://doi.org/10.1201/9781315139470} {\path{doi:10.1201/9781315139470}}.
\newline\urlprefix\url{https://www.taylorfrancis.com/books/9781351460491}

\bibitem{wald_tests_1943}
A.~Wald, \href{https://www.ams.org/tran/1943-054-03/S0002-9947-1943-0012401-3/}{Tests of statistical hypotheses concerning several parameters when the number of observations is large}, Transactions of the American Mathematical Society 54~(3) (1943) 426--482.
\newblock \href {https://doi.org/10.1090/S0002-9947-1943-0012401-3} {\path{doi:10.1090/S0002-9947-1943-0012401-3}}.
\newline\urlprefix\url{https://www.ams.org/tran/1943-054-03/S0002-9947-1943-0012401-3/}

\bibitem{dabernig_simultaneous_2017}
M.~Dabernig, G.~J. Mayr, J.~W. Messner, A.~Zeileis, \href{https://journals.ametsoc.org/doi/10.1175/MWR-D-16-0413.1}{Simultaneous ensemble postprocessing for multiple lead times with standardized anomalies}, Monthly Weather Review 145~(7) (2017) 2523--2531.
\newblock \href {https://doi.org/10.1175/MWR-D-16-0413.1} {\path{doi:10.1175/MWR-D-16-0413.1}}.
\newline\urlprefix\url{https://journals.ametsoc.org/doi/10.1175/MWR-D-16-0413.1}

\bibitem{giacomet_distributional_2023}
C.~L. Giacomet, A.~C.~V. Ramos, H.~S.~D. Moura, T.~Z. Berra, Y.~M. Alves, F.~M. Delpino, et~al., \href{https://archpublichealth.biomedcentral.com/articles/10.1186/s13690-023-01147-7}{A distributional regression approach to modeling the impact of structural and intermediary social determinants on communities burdened by tuberculosis in eastern {Amazonia} – {Brazil}}, Archives of Public Health 81~(1) (2023) 135.
\newblock \href {https://doi.org/10.1186/s13690-023-01147-7} {\path{doi:10.1186/s13690-023-01147-7}}.
\newline\urlprefix\url{https://archpublichealth.biomedcentral.com/articles/10.1186/s13690-023-01147-7}

\bibitem{friederichs_forecast_2012}
P.~Friederichs, T.~L. Thorarinsdottir, \href{https://onlinelibrary.wiley.com/doi/10.1002/env.2176}{Forecast verification for extreme value distributions with an application to probabilistic peak wind prediction}, Environmetrics 23~(7) (2012) 579--594.
\newblock \href {https://doi.org/10.1002/env.2176} {\path{doi:10.1002/env.2176}}.
\newline\urlprefix\url{https://onlinelibrary.wiley.com/doi/10.1002/env.2176}

\bibitem{johnson_quantile-forest_2024}
R.~A. Johnson, \href{https://joss.theoj.org/papers/10.21105/joss.05976}{quantile-forest: a {Python} package for {QuantileRegression} {Forests}}, Journal of Open Source Software 9~(93) (2024) 5976.
\newblock \href {https://doi.org/10.21105/joss.05976} {\path{doi:10.21105/joss.05976}}.
\newline\urlprefix\url{https://joss.theoj.org/papers/10.21105/joss.05976}

\bibitem{mcneil_quantitative_2015}
A.~J. McNeil, R.~Frey, P.~Embrechts, Quantitative risk management: concepts, techniques and tools, revised edition Edition, Princeton {Series} in {Finance}, Princeton University Press, Princeton; Oxford, 2015.

\bibitem{pjm_interconnection_pjm_2021}
L.~PJM~Interconnection, \href{https://www.pjm.com/-/media/DotCom/documents/manuals/archive/m11/m11v112-energy-and-ancillary-services-market-operations-01-05-2021.pdf}{{PJM} manual 11: energy and ancillary services market operations}, Tech. rep., PJM Interconnection, L.L.C., Valley Forge, PA (Jan. 2021).
\newline\urlprefix\url{https://www.pjm.com/-/media/DotCom/documents/manuals/archive/m11/m11v112-energy-and-ancillary-services-market-operations-01-05-2021.pdf}

\bibitem{pjm_interconnection_pjm_2025}
L.~PJM~Interconnection, \href{https://dataminer2.pjm.com}{{PJM} data miner 2}, place: Valley Forge, PA Publisher: PJM Interconnection, L.L.C. (Sep. 2025).
\newline\urlprefix\url{https://dataminer2.pjm.com}

\bibitem{pjm_interconnection_manual_2025}
L.~PJM~Interconnection, {Revision 95}, \href{https://www.pjm.com/-/media/DotCom/documents/manuals/m13.pdf}{Manual 13: emergency operations}, Tech. rep., PJM Interconnection, L.L.C., Norristown, PA (Feb. 2025).
\newline\urlprefix\url{https://www.pjm.com/-/media/DotCom/documents/manuals/m13.pdf}

\bibitem{mayr_generalized_2012}
A.~Mayr, N.~Fenske, B.~Hofner, T.~Kneib, M.~Schmid, \href{https://doi.org/10.1111/j.1467-9876.2011.01033.x}{Generalized additive models for location, scale and shape for high dimensional data — a flexible approach based on boosting}, Journal of the Royal Statistical Society: Series C (Applied Statistics) 61~(3) (2012) 403--427.
\newblock \href {https://doi.org/10.1111/j.1467-9876.2011.01033.x} {\path{doi:10.1111/j.1467-9876.2011.01033.x}}.
\newline\urlprefix\url{https://doi.org/10.1111/j.1467-9876.2011.01033.x}

\bibitem{nerc_north_american_electric_reliability_corporation_2024_2024}
{NERC (North American Electric Reliability Corporation)}, \href{https://www.nerc.com/pa/RAPA/ra/Reliability%20Assessments%20DL/NERC_Long%20Term%20Reliability%20Assessment_2024.pdf}{2024 {Long}-term reliability assessment}, Tech. rep., NERC, Princeton, NJ (2024).
\newline\urlprefix\url{https://www.nerc.com/pa/RAPA/ra/Reliability%20Assessments%20DL/NERC_Long%20Term%20Reliability%20Assessment_2024.pdf}

\end{thebibliography}
\end{document}